 \documentclass[11pt]{article}

 \pdfoutput=1

\usepackage[english]{babel}

\usepackage[round]{natbib} 

\usepackage{graphicx,siunitx,xspace}
\usepackage{amsmath}
\usepackage{amssymb}
\usepackage{hyperref}
\usepackage{enumitem}
\usepackage{color}
\usepackage{enumitem}
\usepackage{float}
\usepackage{amsfonts,bm}
\usepackage{epstopdf}
\usepackage[mathscr]{euscript}

  \usepackage[normalem]{ulem}
  \usepackage{textcomp}

\usepackage[utf8]{inputenc}
\usepackage[T1]{fontenc}
\usepackage{pifont} 

\usepackage{chemformula}

\usepackage{float}
\usepackage{ifthen}
\begin{document}


\begin{center}
{\bf \Large An English translation of the PhD Thesis of Max 
\citet[][]{Planck_1879}}
\\ \vspace*{2mm}
{\bf \Large ``\underline{\"Uber den zweiten Hauptsatz der mechanischen W\"armetheorie\,}''}
\\ \vspace*{2mm}
{\bf \Large or: ``\underline{About the second law of mechanical heat theory\,}''}
\\ \vspace*{2mm}
{\bf \Large to provide a readable version of the German content.}
\\ \vspace*{2mm}
{\bf \large Translated by Dr.Hab. Pascal Marquet 
}
\\ \vspace*{2mm}
{\bf\color{red} \large Possible contact at: 
    pascalmarquet@yahoo.com 
    Web sites$\,$}\footnote{\color{red}$\:$ Google-sites:
    \url{https://sites.google.com/view/pascal-marquet}
    \\ \hspace*{8mm} ArXiv: 
    \url{https://arxiv.org/find/all/1/all:+AND+pascal+marquet/0/1/0/all/0/1}
    \\ \hspace*{8mm} Research-Gate:
    \url{https://www.researchgate.net/profile/Pascal-Marquet/research}
    }
\\ \vspace*{0mm}
\end{center}

\hspace*{65mm} Version-1 / \today
\vspace*{-6mm}

\bibliographystyle{ametsoc2014}
\bibliography{Book_FAQ_Thetas_arXiv}
\vspace*{0mm}

\noindent{\small
{\it {\color{red}Note at the bottom of several pages:} ``\,Planck, Second Law\,''} }
\vspace*{1mm}

Uncertainties/alternatives in the translation are indicated {\color{red} (in red)} with {\it\color{red} italic terms}, together with some additional footnotes (indicated with {\it\color{red} P. Marquet)}. 
Moreover, I have added some highlight (shown as \dashuline{\,dashing text}), in particular about the \dashuline{integration constant issue for the entropy}.

Do not hesitate to contact me in case of mistakes or any trouble in the English translation from the German text.
\vspace*{-4mm}

  \tableofcontents

\newpage

{\Large\bf\underline{Preface of P. Marquet}}
\vspace*{1mm}

In a paper written 4 years before his death about ``{\it The history of the discovery of the physical quantum of 
action}\,''$\,$\footnote{\color{red}\it$\:$I have translated in English the \underline{Planck (1943) paper} and made it available in arXiv and Zenodo, together with the English translation of the \underline{Habilitation-Thesis report of Max Planck (1880)} / See: Planck, M. (1943): Zur Geschichte der Auffindung des physikalischen Wirkungsquantums. Die Naturwissenschaften, {\bf 31} (14), p.153-159, doi:10.1007/BF01475738, German copy available at: \url{https://link.springer.com/content/pdf/10.1007/BF01475738.pdf}}
Max Planck explained that: 
``{\it What has always interested
me most in physics were the great general laws that have significance for all natural processes, regardless
of the properties of the bodies involved in the processes and of the ideas one forms about their structure. I was therefore particularly fascinated by the two laws of thermodynamics. However, while the first law, the law of conservation of energy, has a very simple and easily comprehensible meaning and
therefore requires no special explanation, the correct \dashuline{\,understanding of the second law\,} requires detailed
study. I learnt this principle (second law) in my last year of study (1878) by reading the writings of R.
Clausius (1865-1876), which particularly attracted me anyway because of the excellent clarity and persuasiveness of
the language.\,}''

Max Planck wrote in 1879 at the age of $21$ his ``{\it Munich doctoral dissertation\:}'' (the first part of his German PhD, here translated into English, before his Habilitation in 1880) about the ``{\it interpretation of this second law of thermodynamics\:}'' and starting from the book of Clausius (1865), and in particular for clarifying the differences between ``{\it neutral\:}'' {\it\color{red}(indifferent)} and ``{\it natural\;}'' {\it\color{red}(privileged)} processes.

Max Planck explained that he discovered interesting theorems, but that:
``{\it Unfortunately, however, as I only discovered later, the great American theorist John Willard Gibbs (...) had formulated the same theorems earlier, even in a more general version\,(...)\:}''

Max Planck also confessed that: ``{\it The impression this (Doctoral dissertation) paper had on the physical public at the time was zero.
As I know from conversations with my \dashuline{\,university teachers\,}, none of them had any understanding of its content. They probably only let it pass as a dissertation because they knew me from my other work in the practical course in physics and in the mathematics seminar. But even among physicists who were closer to the subject, I found no interest, let alone applause. 
\dashuline{\,Helmholtz\,} had probably not read the paper at all, and \dashuline{\,Kirchhoff\,} explicitly rejected its content, stating that the concept of entropy, whose magnitude can only be measured and therefore defined by a reversible process, should not be applied to irreversible processes. 
I was unable to get close to \dashuline{\,Clausius\,}, who was very reserved in his personal relationship. An attempt I made once to introduce myself to him in Bonn was unsuccessful, because I
didn't meet him at home.\,}''

However, then Max Planck (1897-1901)  
``{\it discovered new territory in the area of radiant heat\:}'' (...) which led him to his famous \dashuline{\,black-body radiation law\,}.

Indeed, Max Planck explained that:
``{\it In my in-depth study of this problem, destiny made that an external circumstance that I had previously found unpleasant (i.e. the lack of interest among my fellow experts in the research direction I was pursuing) now came as a certain relief to my work. 
At that time, a number of outstanding physicists had turned their attention to the problem of energy distribution in the normal spectrum, both experimentally and theoretically. 
But all of them were only looking in the direction of representing the radiation intensity as a function of the temperature, while \dashuline{\,I suspected the deeper connection in the dependence of the entropy on the energy\,}. 
Since the importance of the concept of entropy had not yet been recognised at that time, nobody cared about the method I used, and I was able to carry out my calculations with leisure and thoroughness without having to fear any interference or revision from any side\,}.

Therefore, the present English version of the German thesis of Planck described the first step toward his deep understanding of thermodynamics, and especially of the concept of entropy, which he will use to arrive at the concept of discrete quanta of exchange of energy in order to compute the Planck-Boltzmann entropy relationship $S=k\:\ln(W)$, and at last to express the Planck's black-body radiation law.
\vspace*{2mm}

\setcounter{section}{-1}
\section{\underline{Introduction {\color{red}of the Thesis}} (p.1-2)} 
\vspace*{-2mm}

While the first law of the mechanical theory of heat, which states the equivalence of heat and work, is based on the principle of the conservation of force and thus deals with a quantity that remains constant in all processes of nature, the second law, in contrast, represents a law according to which nature endeavours to carry out its processes only in a certain sense, in a certain direction, so that it is impossible for the world to return to a previously existing state.
To fix mathematically the sense of this direction --in which all changes in nature take place-- is the meaning of the second law in its most general form.

R. Clausius, the ingenious founder of modern mechanical heat theory, gave this theorem various forms over the course of time, initially only considering so-called cyclic processes, but later also any processes, thereby arriving at the most general form of the theorem.  
His investigations are based on the assumption that he distinguishes between different types of transformations in natural processes {\it\color{red}(here: occurring in the Nature, and not only the ``\,privileged\,'' processes)}, namely the transformation of work into heat and vice versa, the transition of a quantity of heat from one temperature into another, and finally also the change of organisation (``{\color{red}\it Disgregation\,}'').
His argument is based on the principle he first established and is now generally recognized, namely that heat cannot pass by itself from a colder to a warmer body, i.e. without compensation.

However, there is another, as it seems to me, both more general and more direct way, which in particular also makes the somewhat cumbersome distinction between different types of transformations unnecessary, in order to arrive at the \dashuline{\,derivation of the second law of the mechanical theory of heat}, namely \dashuline{\,in its most general form}, from which the more specific conclusions follow easily.

This derivation, along with some interesting applications, forms the first section of the following treatise, while the second section is devoted to a \dashuline{\,discussion of the equivalence values} introduced by Clausius for individual transformations, the meaning of which does not seem to me to extend to transformations of finite size.

After all, it is necessary to abandon a view put forward by a scholar like Clausius only after the most careful examination.
It is largely thanks to his classic treatises that the mechanical theory of heat has gained momentum in recent times.
\vspace*{-2mm}

\section{\underline{First Section {\color{red}of the Thesis}} (p.3-43)}
\vspace*{-2mm}

The following discussions refer to all {\it natural} processes {\it\color{red}(again: occurring in the Nature, and not only the privileged processes?)}, not just in the area of heat.

Let us consider any process in nature that transforms a series of bodies from a certain initial state to a certain final state (whereby all bodies that undergo any changes as a result of the process are always thought of as being included in this).
If we now think of this final state as the initial state of another process, two cases are possible: either this new process can be set up in some way so that it brings about exactly the former initial state, or this cannot be achieved.

The occurrence of one or the other case obviously depends only on the nature of the initial and final states of the first process considered.
Therefore we can distinguish the two possible cases by the following descriptive designation: in the first case nature has an equal preference for the initial and final states; in the second case nature has more preference for the final state than for the initial state; because in the former case a transition is possible in nature between the two states in both directions, but in the latter only in one particular direction.

All 
processes {\it\color{red}occurring in} nature can be categorised according to this distinction: 
\vspace*{-2mm}
\begin{enumerate}
\vspace*{0mm}
\item
\!\!\!) 
those for whose final state nature has the \dashuline{same preference} as for the initial state, and which we would like to call \dashuline{\,{\it neutral} {(\color{red}\it in fact ``\,indifferent\,''\,)} 
processes};
\vspace*{-1mm}
\item
\!\!\!) 
and those for whose final state nature has \dashuline{more preference} than for the initial state, {\it\color{red}(and which we call)} \dashuline{\,{\it natural} 
{(\color{red}\it in fact ``\,privileged\,''\,)} 
processes}.
\end{enumerate}


Neutral {\it\color{red}(indifferent)} processes also include reversible processes, i.e. those processes that can be reversed in the same way, such as the expansion of a gas without external heat input by overcoming a pressure equal to the expansive force of the gas.
An example of a neutral {\it\color{red}(indifferent)} process that cannot be directly reversed is the movement of a freely falling body without friction.
This process can be reversed by forcing the body, after it has fallen freely for a distance, 
  to follow a fixed curve that continuously curves upwards again.
Then the body (assuming no friction) will reach its initial resting position again at a speed of $0$ and thus exactly the former initial state will be restored.

It should not seem superfluous at this point to emphasise a few sentences that follow directly from the above definitions and will be of importance for later considerations:
\vspace*{-2mm}
\begin{enumerate}
\vspace*{-1mm}
\item
\!\!\!) 
 a process cannot be natural {\it\color{red}(privileged)} and neutral {\it\color{red}(indifferent)} at the same time;
\vspace*{-1mm}
\item
\!\!\!) 
a natural {\it\color{red}(privileged)} process can be reversed neither by a natural {\it\color{red}(privileged)} nor by a neutral {\it\color{red}(indifferent)} process, a neutral {\it\color{red}(indifferent)} process can only be reversed by a neutral {\it\color{red}(indifferent)} process, and that always;
\vspace*{-1mm}
\item
\!\!\!) 
if a neutral {\it\color{red}(indifferent)} process is broken down into a series of successive individual processes, each of these individual processes forms a neutral {\it\color{red}(indifferent)} process in its own right;
\vspace*{-1mm}
\item
\!\!\!) 
if a neutral {\it\color{red}(indifferent)} process is carried out after the end of a process, the overall process thus completed has the nature of the first process, i.e. depending on whether this first process was natural {\it\color{red}(privileged)} or neutral {\it\color{red}(indifferent)}, it is also the overall process.
\end{enumerate}

We will now look at the essential properties and characteristics of a natural {\it\color{red}(privileged)}  and a neutral {\it\color{red}(indifferent)} process.
Since these depend only on the nature of the initial and final states, namely on the greater or lesser preference of nature for these states, we are led to the (for the time being only hypothetical) assumption that:
\vspace*{-4mm}
\begin{quote}
{\bf for each given state of a system of bodies there exists a certain function, depending on the determinants of this state, the value of which forms the measure of nature's preference for this state.}
\end{quote}
\vspace*{2mm}

We will denote this function by $S$.
If we now imagine that, with the system of bodies under consideration, some neutral {\it\color{red}(indifferent)} process is carried out from the given state, which puts the bodies into another state to which the function value $S'$ corresponds, we have, according to the definition of $S$ and that of a neutral {\it\color{red}(indifferent)} process: 
\vspace*{-3mm}
\begin{align}
S' \:-\: S \; = \; 0 \: .
\label{Eq_1}
\end{align}
If the process carried out was a natural {\it\color{red}(privileged)} process, then 
\begin{align}
S' \:-\: S \; > \; 0 \: .
\label{Eq_2}
\end{align}
For infinitely small processes, we obtain: 
\vspace*{-3mm}
\begin{align}
dS \; > \; 0 \: 
\nonumber
\end{align}
as properties of a natural {\it\color{red}(privileged)} and a neutral {\it\color{red}(indifferent)} elementary process.

In the same way we say the other way round: if two different states of a system of bodies with the corresponding function values $S$ and $S'$ are given, the value of the quantity $(S'-S)$ decides whether a transition from the first to the second state, or from the second to the first, or between both states in both directions, is possible in nature.
If $(S'-S)$ is positive, we have the first case, if $(S'-S)$ is negative, we have the second case, and if $S'-S=0$, we have the third case.
The first two cases correspond to natural {\it\color{red}(privileged)} processes, the last to neutral {\it\color{red}(indifferent)} processes. 

If the function $S$ is generally known, it can always be used to solve the problem of deciding, given any two given states, which of the two nature has more preference for, i.e. in which direction a transition from one to the other occurs in nature is possible.

Of course, the two given from one to the other must not be associated with either loss of matter or loss of power (in the corresponding sense of this word); rather: 
\vspace*{-2mm}
\begin{enumerate}
\vspace*{-1mm}
\item
\!\!\!) 
both states must contain the same substances, in equal quantities; 
\vspace*{-1mm}
\item
\!\!\!) 
the sum of all 
living forces$\,$\footnote{$\,${\it\color{red}Here, Max Planck retained the old terminology coming from the old Latin term ``\,Vis viva\,'' coined by Gottfried Wilhelm Leibniz and after the work of Christian Huygens, or equivalently  ``\,force vive\,'' in French, or ``\,living force\,'' in English, or ``\,lebendigen Kr\"afte\,'' in German.
Technically, the ``\,Vis viva\,'' $(m\:v^2)$ was twice what is nowadays called the ``\,kinetic energy\,'' $(m\:v^2/2)$ after the result of the work of Gaspard-Gustave Coriolis and Jean-Victor Poncelet.
The ``\,lebendigen Kr\"afte\,'' (living force) of Max Planck is therefore nothing else than the ``\,kinetic energy\,'' $(m\:v^2/2)$, in opposition with the other kind of ``\,dead force\,'' or ``\,potential energy\,'' in English. 
 (P. Marquet)}.}
{\it\color{red} (kinetic energy)} and work supplied (in the most general sense of the word) must be the same in both states.
\end{enumerate}

If these two conditions are not met, then according to the laws of nature a transition from one state to the other is not at all conceivable, and the question of the greater or lesser preference of nature for one or the other state is, in itself, invalid.

We can summarize the results of the last considerations in the following principle:
\vspace*{-4mm}
\begin{quote}
{\bf A transition is always possible in nature between two states with the corresponding function values $S$ and $S'$, which include the same supply of matter and force in both directions, if $S'-S=0$ (by a neutral {\it\color{red}--indifferent--} process), only from the first state to the second if $S'-S>0$ (by a natural {\it\color{red}--privileged--} process), only from the second to the first if $S'-S<0$ (through a natural {\it\color{red}--privileged--} process).}
\end{quote}
\vspace*{-3mm}

The function values $S'$ and $S$ naturally extend to all bodies that are in somehow different conditions in the two states, while all those bodies that behave completely the same in both states can be ignored.

We now want to turn specifically to the area of heat and will prove that if we limit ourselves to processes that belong to this area, the function $S$ really exists and is nothing other than the one in the mechanical theory of heat under the function known as \underline{\it entropy}, as first introduced by 
Clausius.$\,$\footnote{$\,${\it\color{red} Note that the paper of Clausius  (1865), where he called with the name ``\,Entropie\,'' the state function $S$ he defined with the relationship $dS=\int dQ/T$, had been published only 14 years before the Thesis of Planck (1879). 
Note also that Planck did not cite in 1879 the works of Boltzmann, because it was only two years after Boltzmann defined in 1877 (in a long and complex contribution still unknown in 1879) the ``\,Entropie\,'' of a macroscopic state in terms of ``\,the logarithm of the number of complexions that realize that state\,'' and next leading to the definition $S=k\:\ln(W)+cste$ that Planck will derive in 1900-1901 to arrive at his famous relationship for the  black-body radiation. (P. Marquet)}.}
At the same time, the application of the principles expressed above will give us the second law of the mechanical theory of heat in its most general form.

The way we have to proceed, in order to find the characteristics and properties of a natural {\it\color{red}(privileged)} and a neutral {\it\color{red}(indifferent)} process in the field of heat, can be characterized as follows: we consider two different states from several bodies, which, however, have the same supply of matter and force contained, and first look for the sufficient condition that a neutral {\it\color{red}(indifferent)} process is possible between them, then the sufficient condition that a natural {\it\color{red}(privileged)} process is possible from the first to the second state.
The results will then also enable us to specify the necessary conditions for the above cases, i.e. the equations that must apply if the transition from the first to the second can be accomplished by a neutral {\it\color{red}(indifferent)} or a natural {\it\color{red}(privileged)} process.

First of all, we will only consider those bodies in the circle of investigations in which no internal work is done during the changes of state under consideration, whose heat capacity is constant, and which exactly follow the law of Mariotte and Gay-Lussac, i.e. the so-called perfect gases; because for these perfect gases the expression of entropy results immediately, while for all other bodies the existence of this function must first be specifically proven.

\subsection{\underline{Perfect gases} (p.8-28)}
\vspace*{-1mm}

We initially limit ourselves to comparing only those states that differ infinitely from each other, so that the results can then be easily generalized.
\vspace*{2mm}

\subsubsection{\underline{Comparison of two states of a single gas} (p.8-12)}
\vspace*{-2mm}

The true heat capacity of the gas, based on its mass, is $c$, and the constant of Gay-Lussac and Marriotte's law is 
$$ k \;=\; \frac{v\:.\:p}{T} \;\;\; \mbox{($v$ volume, $p$ pressure, $T$ absolute temperature) \:.}$$ 
We adopt $T$ and $v$ for independent variables. Finally, it should be noted that all occurring quantities of heat should be measured according to mechanical measures {\it\color{red} (at that time: $1$~``calories'' $\approx4.18$~J and $1$~``erg'' $=10^{-7}$~J for the energies)}.

Now let us take as given any two infinitely different states of the gas, which are characterized by the corresponding values of temperature and volume:
\vspace*{1mm}\\
\hspace*{45mm} 1st state / $T$, $v$ ; 
\vspace*{1mm}\\
\hspace*{45mm} 2nd state / $T+dT$, $v+dv$ .
\vspace*{1mm}\\
Since, according to what we have seen and in order to be fully comparable, both states must contain not only the same content of matter but also the same supply of force, and since the energy of the gas in the first state is $c\:.\:T$ and in the second $c\:.\:(T+dT)$, we have to add a (positive or negative) amount of work $(=c\:.\:dT)$ to the first state, which we think of as mechanical work.

We can for instance think of both states forming a heavy body which, under otherwise identical circumstances, is at a different height in the first state than in the second, in such a way that the difference between the mechanical forces present in the body in both states is the previous value of work $c\:.\:dT$.
Or we can also assume that the heavy body is at rest in one state, but is endowed in the other (at the same height) with a certain living force {\it\color{red} (kinetic energy)}, which corresponds to the above value of work.
If we make one of these assumptions, both states are provided with equal supplies of matter and force, and are therefore completely comparable.

In order to avoid any misunderstandings, a general remark may be made here, which also applies to many of the following considerations: if we are dealing with the investigation of a process which brings a series of bodies from one particular state into another, it is completely irrelevant to the nature of this process whether such bodies (which are not thought to be included in the two states) also suffer temporary changes as a result of it
 (only if, again, they behave exactly the same in the final state as in the initial state).
Because the condition is only that the process, starting from the initial state, brings about exactly the final state, and no conditional presupposition is made about the nature of the transition.
The nature of the process depends only on the determinants {\it\color{red}(i.e. physical properties)} of the initial and final states, and here all those bodies which are in exactly the same conditions in both states can be disregarded.

In order to find the sufficient condition for the possibility of a neutral {\it\color{red}(indifferent)} process between the two given states, we first imagine the gas being brought from the first state to the volume $v+dv$ required for the second state by a reversible process, i.e. by overcoming a pressure equal to the expansive force of the gas, and without the addition of heat from outside, whereby the temperature may change to $T+\Delta \: T$.
Then we have the equation: 
$$ 0 \; = \; c \:.\;\Delta T \:+\: p \:.\; dv \: .$$
Since $p = k \:.\: T / v$, we now have: 
\vspace*{-3mm}
\begin{align}
0 \; = \; c \:.\;\Delta T \:+\: \frac{k\:T}{v} \; dv \: .
\label{Eq_3}
\end{align}
The resulting state ($T+\Delta T$, $v+\Delta v$) can in any case be transformed into the state ($T+ d T$, $v+ d v$) by a neutral {\it\color{red}(indifferent)} process if $\Delta T = dT$, because in this case the gas is already in its prescribed final state and one then only needs to allow the mechanical working variables to merge into one another, which can always be accomplished by a neutral {\it\color{red}(indifferent)} process.

By substituting $dT$ instead of $\Delta T$ in equation (\ref{Eq_3}), a sufficient condition for a neutral {\it\color{red}(indifferent)} process between the two given states is thus obtained: 
\vspace*{-2mm}
$$ c \:.\; d T \:+\: \frac{k\:.\:T}{v} \:.\; dv 
\; = \; 0 \: ,$$
or: 
$$ c \:.\; \frac{d T}{T} \:+\: k \:.\; \frac{d v}{v} 
\; = \; 0 \: ,$$
i.e. 
\vspace*{-2mm}
\begin{align}
dS \; = \; 0 \: 
\label{Eq_4}
\end{align}
if $S$ denotes the entropy of the gas
$$ 
S \; = \; c \; \ln(T) \:+\: k \; \ln(v) 
\:+\: const.  $$
This is the condition whose fulfilment makes a neutral {\it\color{red}(indifferent)} process between the two states possible in any case, because if such a process is possible from the first to the second state, one can also be carried out in the opposite direction according to the definition of a neutral {\it\color{red}(indifferent)} 
process.$\,$\footnote{$\:${\it\color{red} The result $dS=0$ derived in (\ref{Eq_4}), together with the status of processes ``\,possible from the first to the second state'', show that this ``\,neutral\,'' processes of Planck correspond to ``\,adiabatic and reversible\,'', and thus ``\,isentropic'' processes (P. Marquet)}.}

Let us now move on to establishing the sufficient condition for the possibility of a natural {\it\color{red}(privileged)} process from the first to the second state.
A natural {\it\color{red}(privileged)} process occurs when a perfect gas expands without overcoming an external pressure, i.e. at a constant 
temperature.$\,$\footnote{\label{note_p10}$\:${A note from Planck: 
In other words, a perfect gas cannot reduce its volume at a constant temperature without compensation.
This principle, generalised to any body, is: ``A body cannot reduce its volume without compensation at constant energy'', and is quite analogous to Clausius' well-known principle of heat transfer from a colder to a warmer body, which cannot take place without compensation.
Both principles are mutually dependent, but I have preferred to use the former because it only deals with a single body. 
}}
So if $dv$ was positive and $dT=0$, the transition from the first to the second state could in any case be achieved by a natural {\it\color{red}(privileged)} process.
If we take these two values (i.e. $dv>0$ and $dT=0$) as the given ones and insert them into the expression of $dS$: 
$$ c \:.\; \frac{d T}{T} \:+\: k \:.\; \frac{d v}{v} $$ 
we get: 
$$dS \; = \; k \:.\; \frac{d v}{v} \; > \; 0 \: , $$ 
an essentially positive expression under the condition made, which is why we have:
\begin{align}
dS \; > \; 0  \: 
\label{Eq_5}
\end{align}
for this case.

From this, however, it can be directly concluded that this equation (\ref{Eq_5}) also generally expresses the sufficient condition for the possibility of a natural {\it\color{red}(privileged)} process from the first to the second state.
Because if the two given states satisfy this equation, we can always use a natural {\it\color{red}(privileged)} process of the type considered above (by allowing the gas to expand properly without overcoming external pressure) to bring about a state in which the entropy of the gas $S + dS$ is that of the second state, and then bring the gas exactly into the second given state by a neutral {\it\color{red}(indifferent)} process while maintaining this value of entropy. Then the transfer from the first to the second state is accomplished entirely through a natural {\it\color{red}(privileged)} process.

The mechanical working variables that are ultimately to be converted into one another always require a neutral {\it\color{red}(indifferent)} process and therefore have no influence on the nature of the entire process. We can therefore ignore them here, as well as later in similar considerations.

Likewise, $dS<0$ is the sufficient condition for the possibility of a natural {\it\color{red}(privileged)} process from the second to the first state.

From this it finally follows that $dS>0$ and $dS=0$ are at the same time the necessary conditions that must be fulfilled if a {\it natural} {\it\color{red}(privileged)} or {\it neutral} {\it\color{red}(indifferent)} process from the first to the second state is to be possible, respectively.

Indeed, if we first imagine that the second state emerged from the first through some natural {\it\color{red}(privileged)} process, then $dS$ cannot be $=0$, because then, according to equation (\ref{Eq_4}), a neutral {\it\color{red}(indifferent)} process would be possible that brought about the same result, which contradicts the assumption since, according to the definitions of natural {\it\color{red}(privileged)} and neutral {\it\color{red}(indifferent)} processes, only one type of process is possible between two specific states.
Furthermore, $dS$ cannot be $<0$ either, because then a natural {\it\color{red}(privileged)} process from the second to the first state would be possible, which is also incompatible with the assumption made. 
Consequently, $dS>0$ must be.

In a very similar way it can be shown that $dS=0$ is the necessary condition of a neutral {\it\color{red}(indifferent)} process from the first to the second state.
\vspace*{0mm}

\subsubsection{\underline{Comparison of two states of two gases} (p.12-15)}
\vspace*{-1mm}

The two given states are characterized as follows: 
\vspace*{2mm}\\ \hspace*{15mm} State (1):
$$
\mbox{temperature of the 1st gas: }\;T_1\:,
\mbox{volume}\;v_1\:,
\mbox{entropie}\;S_1\:;
$$
$$
\mbox{temperature of the 2nd gas: }\;T_2\:,
\mbox{volume}\;v_2\:,
\mbox{entropie}\;S_2\:;
$$
\vspace*{-4mm}\\ \hspace*{15mm} State (2):
$$
\mbox{entropie of the 1st gas: }\;S_1\:+\:dS_1\:;
$$
$$
\mbox{entropie of the 2nd gas: }\;S_2\:+\:dS_2\:.
$$
Here again, so that the two states contain equal reserves of power, we must think of one of them as having a corresponding mechanical work reserve, equal to the difference in the total energies of the two states. Then the two states are completely comparable.

When examining the characteristics and properties of a process that, starting from the first state, brings about the second, it is obviously permissible to substitute any other state instead of the second state, which is characterized by the same entropies $S_1+dS_1$ and $S_2+dS_2$ of the two gases (with the corresponding mechanical work supply).
Because when the transition from the first to this inserted state has been made, each gas can be brought into the second state by itself through a reversible process in which its entropy remains constant, and the nature of the entire process is not changed by this {\it\color{red}if} you add a neutral {\it\color{red}(indifferent)} process at the end.
Therefore we only have to provide the values of the entropies to characterize the second state.

In order to establish the sufficient condition for the possibility of a neutral {\it\color{red}(indifferent)} process, we imagine that a neutral {\it\color{red}(indifferent)} process is actually carried out from the first state, which should ultimately bring about exactly the second given, or the substituted state mentioned, i.e. bring the two gases to the entropies $S_1+dS_1$ and $S_2+dS_2$.
This can be done in the following way: we first bring both gases to the same temperature $T$ by compressing each of them individually in a reversible manner without adding heat from the outside and then put them in such a connection with each other that they can communicate heat to each other but cannot exchange their pressure. 
This is of course perfectly conceivable and has already been used repeatedly as evidence.
Because the temperatures are the same, the gases will not initially share any heat.
But now let the first gas, for example, expand or contract in a reversible manner, then the temperatures in both gases will equalise at any moment, and in a reversible manner.
Since the first gas absorbs or releases heat from the outside, its entropy will change and it is possible to assume the value $S_1+dS_1$.
As soon as this has occurred, isolate both gases again, whereby it may be found that the second gas has the assumed entropy $S_2+\Delta\,S_2$.

In order to be able to reach the second given state by a neutral {\it\color{red}(indifferent)} process, obviously only the {\it\color{red}following} condition must be fulfilled: 
\vspace*{-2mm}
\begin{align}
dS_2 \; = \; \Delta S_2 \: .
\label{Eq_6}
\end{align}

The value of $\Delta S_2$ can be easily specified: at the moment when the two gases with the common temperature $T$ were put into thermally conductive contact, their entropies were still $S_1$ and $S_2$.
If the amount of heat $dQ$ has now subsequently been transferred from the second to the first gas, causing the entropies to change by $dS_1$ and $\Delta\,S_2$ respectively, then, since this heat exchange took place in a reversible manner: 
\vspace*{1mm}\\
\hspace*{35mm} 
For the 1st gas: $\;dQ=T \:.\; dS_1\:,$ 
\vspace*{1mm}\\
\hspace*{35mm} 
For the 2nd gas: $\;-\,dQ=T \:.\; \Delta S_2\:,$
\vspace*{1mm}\\ 
from which by addition: 
\begin{align}
dS_1 \;+\; \Delta S_2 \; = \; 0 \: .
\label{Eq_7}
\end{align}
If we now introduce the condition of equation (\ref{Eq_6}) here, 
\begin{align}
\boxed{\:dS_1 \;+\; dS_2 \; = \; 0 \:} 
\label{Eq_8}
\end{align}
results as a \dashuline{\,sufficient condition\:} for the possibility of a  \dashuline{\,neutral\;} {\it\color{red}(indifferent)}  process between the two states.

If equation (\ref{Eq_6}) is not satisfied, but if e.g.  
\begin{align}
dS_2 \; > \; \Delta S_2  \: ,
\label{Eq_9}
\end{align}
then the transition to the second state could occur through a natural {\it\color{red}(privileged)} process.
Because then, after everything has been done in the manner described, one could finally bring the second gas on its own through a natural {\it\color{red}(privileged)} process from the entropy $S_2+\Delta S_2$ to the entropy $S_2+ d S_2$, since the one previously derived for this purpose is the condition (\ref{Eq_5}), which applied to this case leads to $d S_2 - \Delta S_2 > 0$, is then fulfilled by the assumption (\ref{Eq_9}).
If we now eliminate $\Delta S_2$ from the always valid equation (\ref{Eq_7}) and from (\ref{Eq_9}), then 
\vspace*{-3mm}
\begin{align}
\boxed{\:dS_1 \; + \; dS_2 \; > \; 0 \:}
\label{Eq_10}
\end{align}
gives us a \dashuline{\,sufficient condition\:} for the possibility of a \dashuline{\,natural\;} {\it\color{red}(privileged)} process from the first to the second state.

Conversely, $dS_1+dS_2<0$ allows a natural {\it\color{red}(privileged)} process from the second to the first state.

These sufficient conditions finally turn out to be the necessary ones.
Because if we e.g. assuming that the second state emerged from the first through a natural {\it\color{red}(privileged)} process, then: 1) $dS_1+dS_2$ cannot be $=0$, because otherwise a neutral {\it\color{red}(indifferent)} process would be possible\,; 2) $dS_1+dS_2$ cannot $ <0$ because otherwise from the second to the first state would be a natural {\it\color{red}(privileged)} process.
Consequently, $dS_1+dS_2$ must be $> 0$.
Likewise, the equation {\it\color{red}(\ref{Eq_10})} 
$\boxed{\:dS_1 \:+\: dS_2 \;=\; 0\:}$ 
{\it\color{red}also} forms the \dashuline{\,necessary condition} for the possibility of a \dashuline{\,neutral\;} {\it\color{red}(indifferent)}  process between the two states.
\vspace*{-2mm}

\subsubsection{\underline{Comparison of two states of three or more gases} (p.15-28)}
\vspace*{-1mm}

It is now easy to show how the results obtained can be extended to the comparison of two states of three or more gases.
Let's first look at $3$ gases:
\vspace*{1mm}\\
\hspace*{25mm}1st state / Entropies: 
\hspace*{11mm}$S_1$ 
\hspace*{16mm}$S_2$ 
\hspace*{16mm}$S_3$
\vspace*{1mm}\\
\hspace*{25mm}2nd state / Entropies: 
\hspace*{5mm}$S_1+dS_1$ 
\hspace*{5mm}$S_2+dS_2$ 
\hspace*{5mm}$S_3+dS_3$
\vspace*{1mm}\\
along with the corresponding mechanical work supply.

To establish the sufficient condition for the possibility of a neutral {\it\color{red}(indifferent)} process between the two given states, we organize a neutral {\it\color{red}(indifferent)} process that, starting from the first state, should bring about exactly the second.
First, we leave the third gas unchanged and imagine a reversible process carried out with only the first two gases in such a way that the entropy of the first gas becomes $S_1+dS_1$, whereby that of the second gas changes to $S_2+\Delta S_2$.
This change is possible according to the above if the following condition exists:
\vspace*{-3mm}
\begin{align}
dS_1 \; + \; \Delta S_2 \; = \; 0 \: .
\label{Eq_11}
\end{align}
Finally, we have to bring the last two gases with the entropies $S_2+\Delta S_2$ and $S_3$ into their final state by a neutral {\it\color{red}(indifferent)} process, to which the entropies $S_2+d S_2$ and $S_3+d S_3$ correspond.
The changes in entropies brought about by this latter process are: 
$$dS_2 \;-\; \Delta S_2 \;\;\;\mbox{and}\;\;\; dS_3 \: ,$$
resulting in the condition: 
\vspace*{-3mm}
\begin{align}
dS_2 \; - \; \Delta S_2 \; + \; dS_3\; = \; 0 \: ,
\label{Eq_12}
\end{align}
with (11) and (12) adding up to: 
\vspace*{-2mm}
\begin{align}
\boxed{\:dS_1 \; + \; d S_2 \; + \; dS_3\; = \; 0 \:}
\label{Eq_13}
\end{align}
as the 
\dashuline{\,sufficient condition\:} for the possibility of a  \dashuline{\,neutral\;} {\it\color{red}(indifferent)} 
process between the two given states.

By complete induction one can easily obtain the corresponding general result for any number of gases.

We first look for the sufficient condition for the possibility of a neutral {\it\color{red}(indifferent)} process between the following 2 states of $n$ different amounts of gas: 
\vspace*{1mm}\\
\hspace*{25mm}1st state / Entropies: 
\hspace*{11mm}$S_1$ 
\hspace*{16mm}$S_2$ 
\hspace*{9mm} ... 
\hspace*{10mm}$S_n$
\vspace*{1mm}\\
\hspace*{25mm}2nd state / Entropies: 
\hspace*{5mm}$S_1+dS_1$ 
\hspace*{5mm}$S_2+dS_2$ 
\hspace*{3mm} ... 
\hspace*{5mm}$S_n+dS_n$
\vspace*{1mm}\\
together with the corresponding mechanical work supply.

In establishing the above condition, we make the assumption that the same condition for $(n-1)$ gas quantities is:
\vspace*{-2mm}
\begin{align}
dS_1 \; + \; d S_2 \; + \; ... \; + \;dS_{n-1}\; = \; 0 \: .
\label{Eq_14}
\end{align}
This assumption gives us, if we leave the $n$-gas unchanged and carry out a reversible process with the $(n-1)$ first gases, so that the entropies are respectively change into  
$S_1+dS_1$, $S_2+dS_2$, ..., 
$S_{n-2}+dS_{n-2}$, $S_{n-1}+\Delta S_{n-1}$,
the following equation as a sufficient condition:
\begin{align}
dS_1 \; + \; d S_2 \; + \; ... \; + \; d S_{n-2}
     \; + \; \Delta S_{n-1}\; = \; 0 \: .
\label{Eq_15}
\end{align}
If we finally bring the last two gases through a reversible process from the entropies 
$S_{n-1}+\Delta S_{n-1}$ and $S_n$ 
to the entropies 
$S_{n-1}+d S_{n-1}$ and $S_n+d S_n$, 
we get as a further condition: 
\begin{align}
dS_{n-1} \; - \; \Delta S_{n-1} \; + \; d S_n \; = \; 0 \: ,
\label{Eq_16}
\end{align}
and by combining the Equations (\ref{Eq_15}) and (\ref{Eq_16}):
\begin{align}
\boxed{\:
dS_1 \; + \; d S_2 \; + \; ... \; + \; d S_n \; = \; 0 \:} 
\: ,
\label{Eq_17}
\end{align}
which is the condition you are looking for.

Now, however, the above condition (\ref{Eq_14}) is proven to be valid for $n=4$, and consequently also for all higher values of $n$.

The sufficient condition for the possibility of a natural {\it\color{red}(privileged)} process between the two (above) states in the direction from the first to the second results directly from this.

If we carry out the same reversible process again with the $(n-1)$ first gas quantities, under the same condition given in equation (\ref{Eq_14}), whereby we also obtain equation (\ref{Eq_15}) again, it is then unnecessary to convert the last two gases with the entropies $S_{n-1}+\Delta S_{n-1}$ and $S_n$ into their final state with the entropies $S_{n-1}+dS_{n-1}$ and $S_n+dS_n$.
However, this can always be achieved by a natural {\it\color{red}(privileged)} process if the sum of the changes of both entropies is 
\begin{align}
dS_{n-1} \; - \; \Delta S_{n-1} \; + \; d S_n \; > \; 0 \: 
\label{Eq_18}
\end{align}
(sufficient condition of a natural {\it\color{red}(privileged)} process with two gases), therefore by uniting (\ref{Eq_15}) and (\ref{Eq_18}) we obtain: 
\vspace*{-3mm}
\begin{align}
dS_1 \; + \; d S_2 \; + \; ... \; + \; d S_n \; > \; 0 \: 
\label{Eq_19}
\end{align}
as the desired condition.

After considerations that are quite analogous to those already made above, it finally results that 
$$\boxed{\:
dS_1 \;+\; dS_2 \;+\; ... \;+\; dS_n \;\geq\; 0
\:}
$$ 
are also the 
\dashuline{\,necessary conditions\:} for the possibility of a \dashuline{\,natural\;} {\it\color{red}(privileged)}  and a \dashuline{\,neutral\;} {\it\color{red}(indifferent)} process
from the first to the second state.

As for such states, which are only infinitely little different from each other, the corresponding theorems also apply to the comparison of any number of perfect gases, which can be easily proven by integrating the obtained conditional equations.

Let us take $2$ different states of $n$ quantities of gas as given, and look for the sufficient condition for a neutral {\it\color{red}(indifferent)} process to be possible between them.

The first state is characterised by the entropies of the gases
\vspace*{2mm}\\
\hspace*{55mm}
        $S_1 \;\; S_2 \;\; ... \;\;  S_n$, 
\vspace*{1mm}\\
the second by the entropies 
\vspace*{1mm}\\
\hspace*{55mm}
        $S'_1 \;\; S'_2 \;\; ... \;\;  S'_n \: ,$ 
\vspace*{1mm}\\
together with the corresponding mechanical energy supply.

If we now, starting from the first state, organise any neutral {\it\color{red}(indifferent)} elementary process, we can, according to the above, give the gases any desired entropy changes, if only the condition (\ref{Eq_17}) is fulfilled: 
\begin{align}
dS_1 \; + \; d S_2 \; + \; ... \; + \; d S_n \; = \; 0 \: .
\nonumber
\end{align}
If we continue the procedure and organise an infinite number of such elementary processes one after the other, we arrive at finite entropy changes by integrating this equation, and at the same time at the condition to which these changes are subject: 
$$
\left( \: S'_1 \; - \; S_1 \: \right)
\; + \;
\left( \: S'_2 \; - \; S_2 \: \right)
\; + \;
. . .
\; + \;
\left( \: S'_n \; - \; S_n \: \right)
\; = \; 0 
$$ 
or: 
\vspace*{-5mm}
\begin{align}
 \boxed{\:\sum \left( \: S' \; - \; S \: \right) \; = \; 0 \:} 
\label{Eq_20}
\end{align}
which is therefore the condition we are looking for.
Similarly, 
\begin{align}
 \boxed{\:\sum \left( \: S' \; - \; S \: \right) \; > \; 0 \:} 
\label{Eq_21}
\end{align}
is the 
\dashuline{\,sufficient conditions\:} for the possibility of a \dashuline{\,natural\;} {\it\color{red}(privileged)} process
from the first to the second state, and finally the necessity of these conditions also follows from the proof of the impossibility of the contrary, 
because if the second state arose from the first through a natural  {\it\color{red}(privileged)} process, then according to the above $\sum \left( \: S' \; - \; S \: \right)$ can neither be $<0$ nor $=0$, etc.

If we call the expression 
$\sum \left( \: S' \; - \; S \: \right)$ 
the entropy value-{\it\color{red} (change)} of the transition from the first to the second state, we can express the result obtained as follows: the entropy value-{\it\color{red} (change)} of the transition decides by its sign whether this transition occurs through a natural {\it\color{red}(privileged)} or a neutral  {\it\color{red}(indifferent)} process or cannot be accomplished at all.
In the first case the entropy value-{\it\color{red} (change)} is positive, in the second $=0$, in the third case negative.
Only in the second case is a transition in both directions possible between the two states.
It should be emphasized again that all those bodies that are under different conditions in both states make contributions to the entropy value-{\it\color{red} (change)}, while those bodies that behave completely the same in both states can be ignored.
Furthermore, the amount of force included in both states in the form of mechanical work or mechanical living force {\it\color{red} (kinetic energy)} also have no influence on the entropy value-{\it\color{red} (change)}.

Conversely, it follows from the theorem obtained that the entropy value-{\it\color{red} (change)} of any process with perfect gases is positive or $=0$\,: in the first case the process is natural {\it\color{red}(privileged)}, in the second neutral {\it\color{red}(indifferent)} and can therefore be reversed.
This characterises a very specific direction in all the effects of nature.

So if you want to bring about a change of state of a group of gases with a negative entropy value-{\it\color{red} (change)}, this can only be done if another group (of gases) undergoes a change of state with a positive entropy value-{\it\color{red} (change)} in connection with it, in such a way that the entropy value of the overall change becomes positive.
One can therefore regard the second change of state as the compensation necessary for the realisation of the first, and must then take as the measure of this compensation that {\it\color{red} (positive)} entropy value-{\it\color{red} (change)} which the second change of state must have so that the entropy value of the total change is at least $=0$, i.e. the absolute value of the entropy value-{\it\color{red} (change)} of the first change of state.

Before we go on to generalize these theorems to any bodies, we want to actually determine the entropy values-{\it\color{red} (change)} for some simple cases and compare the results of the calculation with those of experience in order to prove the complete agreement of the two.

\newpage 
\begin{center}
-------- \underline{First Example (p.20-25)} -------- 
\end{center}
\vspace*{-2mm}

We imagine two perfect gases in arbitrary quantities with the  true-{\it\color{red} (specific)} heat capacities $c_1$ and $c_2$ (based on the masses), the temperatures $T_1$ and $T_2$, and the volumes $v_1$ and $v_2$.

Furthermore, let the constants of Gay-Lussac's and Marriotte's law be $k_1$ and $k_2$, so that $k_1=p_1\:v_1/T_1$ and $k_2=p_2\:v_2/T_2$, where $p$ is the {\it\color{red} (common)}  pressure of gas.

We must determine the entropy value-{\it\color{red} (change)} of the change of state that occurs when the amount of heat $Q$ is transferred from the first gas to the second. 
We will initially assume that the volumes are constant.

To calculate the entropy value-{\it\color{red} (change)}, it is only necessary to know the characteristics of the initial and final states.
Those of the former are $T_1\:v_1$ and $T_2\:v_2$, and those of the latter are $T'_1\:v_1$ and $T'_2\:v_2$. It is therefore not necessary to calculate $T'_1$ and $T'_2$, because  
this is done by the equations : 
\vspace*{1mm} \\ \hspace*{67mm}
$c_1 \: T_1 \:-\: Q \;=\; c_1 \: T'_1$ 
\vspace*{1mm} \\ \hspace*{25mm}
and on the other hand : 
$c_2 \: T_2 \:+\: Q \;=\; c_2 \: T'_2$ 
\vspace*{1mm} \\ 
from which: 
\vspace*{-3mm}
\begin{align}
 T'_1 \; = \; T_1 \; - \; \frac{Q}{c_1}
\label{Eq_22}
\end{align}
\vspace*{-3mm}
and 
\vspace*{-3mm}
\begin{align}
 T'_2 \; = \; T_2 \; + \; \frac{Q}{c_2} \: .
\label{Eq_23}
\end{align}
Consequently, the entropy value-{\it\color{red} (change) is}: 
\begin{align}
 \sum \left( S' \, - \, S \right) 
 & \: = \; 
 \left[\: c_1\:\ln(T'_1) \:+\: k_1\:\ln(v_1) \:\right]
 \;-\;
 \left[\: c_1\:\ln(T_1) \:+\: k_1\:\ln(v_1) \:\right]
\nonumber \\
 & \,\: + \; 
 \left[\: c_2\:\ln(T'_2) \:+\: k_2\:\ln(v_2) \:\right]
 \;-\;
 \left[\: c_2\:\ln(T_2) \:+\: k_2\:\ln(v_2) \:\right]
\nonumber \\
 & \: = \; 
  c_1\:\ln\!\left(\frac{T'_1}{T_1}\right) 
  \;+\;
  c_2\:\ln\!\left(\frac{T'_2}{T_2}\right) 
\nonumber
\end{align}
or, if the values of $T'_1$ and $T'_2$ are used:
\begin{align}
 \sum \left( S' \, - \, S \right) 
 & \: = \; 
  c_1\:\ln\!\left(1 \:-\: \frac{Q}{c_1\:T_1}\right) 
  \;+\;
  c_2\:\ln\!\left(1 \:+\: \frac{Q}{c_2\:T_2}\right) 
\label{Eq_24} \: .
\end{align}
This expression (\ref{Eq_24}) is composed only of given {\it\color{red} (known)} values.
Depending on the value of $Q$, the entropy value-{\it\color{red} (change)} will modify its magnitude.
If (\ref{Eq_24}) is positive the change in state can be carried out through a natural {\it\color{red}(privileged)} process, and if it is $=0$ the transfer takes place through a neutral {\it\color{red}(indifferent)} process, but if it is negative  the change in state cannot take place in nature (unless a compensatory process with a sufficiently large positive entropy value-{\it\color{red} (change)} takes place).

We now want to examine the values-{\it\color{red} (change)} of $\sum \left( S' \, - \, S \right)$ in their dependence on $Q$ and first consider the results of immediate observation.

To fix the ideas, let's consider once and for all that $T_1>T_2$.
Then we can see from experience that nature will act as far as possible to create heat from the first gas into the second, as follows from the laws of heat conduction.

So if $Q$ is negative, i.e. if heat passes from the second gas into the first, the change in state brought about by this will in no way correspond to the meaning of natural {\it\color{red}(privileged)} effects, from which we conclude that the entropy value-{\it\color{red} (change)} $\sum \left( S' \, - \, S \right)$ will then be negative.

The smallest possible value of $Q$ is $-\:c_2\:T_2$, and in that case all the heat from the second gas would transfer into the first.

For $Q=0$ the change of state is vanishing, so it can be carried out by a neutral {\it\color{red}(indifferent)} process, since only such a process can be reversed. It follows that then $\boxed{\:\sum \left( S' \, - \, S \right) = 0\:}$.

Now we assume \underline{$Q$ to be positive}. For small positive values of $Q$, the corresponding change in state can obviously be brought about by a natural {\it\color{red}(privileged)} process, i.e. nature has more preference for the second state than the first, so $\boxed{\:\mbox{$\sum \left( S' \, - \, S \right)$ will be positive}\:}$.
But if $Q$ continues to grow, then in the second state the temperature of the first gas $T'_1$ will always be lower, and that of the second gas $T'_2$ will always be higher, until a certain value $Q \times T'_1 = T_2$ {\it\color{red}is reached}.
Then, according to experience, nature certainly has a greater preference for the second situation.
For even larger values of $Q$, $T'_1$ will become even smaller than $T'_2$, but as long as the temperature difference is {\it\color{red} minimal}, one must assume that nature still retains more preference for the second state than for the first, because it generally strives to compensate for temperature differences as much as possible.
But the moment will come when $Q$ has been 
{\it\color{red}  cancelled / washed so far}, 
and when the difference in temperatures in the second state has become so great that nature's preference will again tend 
towards the first state.
A limit point will be reached where nature's preference for both states will be the same, i.e. $\boxed{\:\sum \left( S' \, - \, S \right) = 0\:}$, and from this limit point onwards, with further growth of $Q$, the preference for the first state will increase, so 
$\boxed{\mbox{\:$\sum \left( S' \, - \, S \right)$ becomes negative}}$.
The maximum value of $Q$ is $c_1\:T_1$, because then all the heat from the first gas would be passed into the second.
If both 
\dashuline{\,gas quanta}$\,$\footnote{$\:${\it\color{red} Note the first appearance of the name ``\,quanta\,'' in the German word  ``\,Gasquanta\,'' p.22 of the PhD text of \cite{Planck_1879}  (P. Marquet)}.} 
are of the same substance and have the same volumes and the same densities, then the limit point at which nature's preference for both states is the same will obviously be reached when $Q$ is so large that the gases have just exchanged their temperatures.
For the general case, however, this type of consideration does not provide a clear result and the latter can only be obtained by calculation.

\begin{figure}[hbt]
\centering
\vspace*{2mm} \hrule  \vspace*{2mm} 
\includegraphics[width=0.49\linewidth]{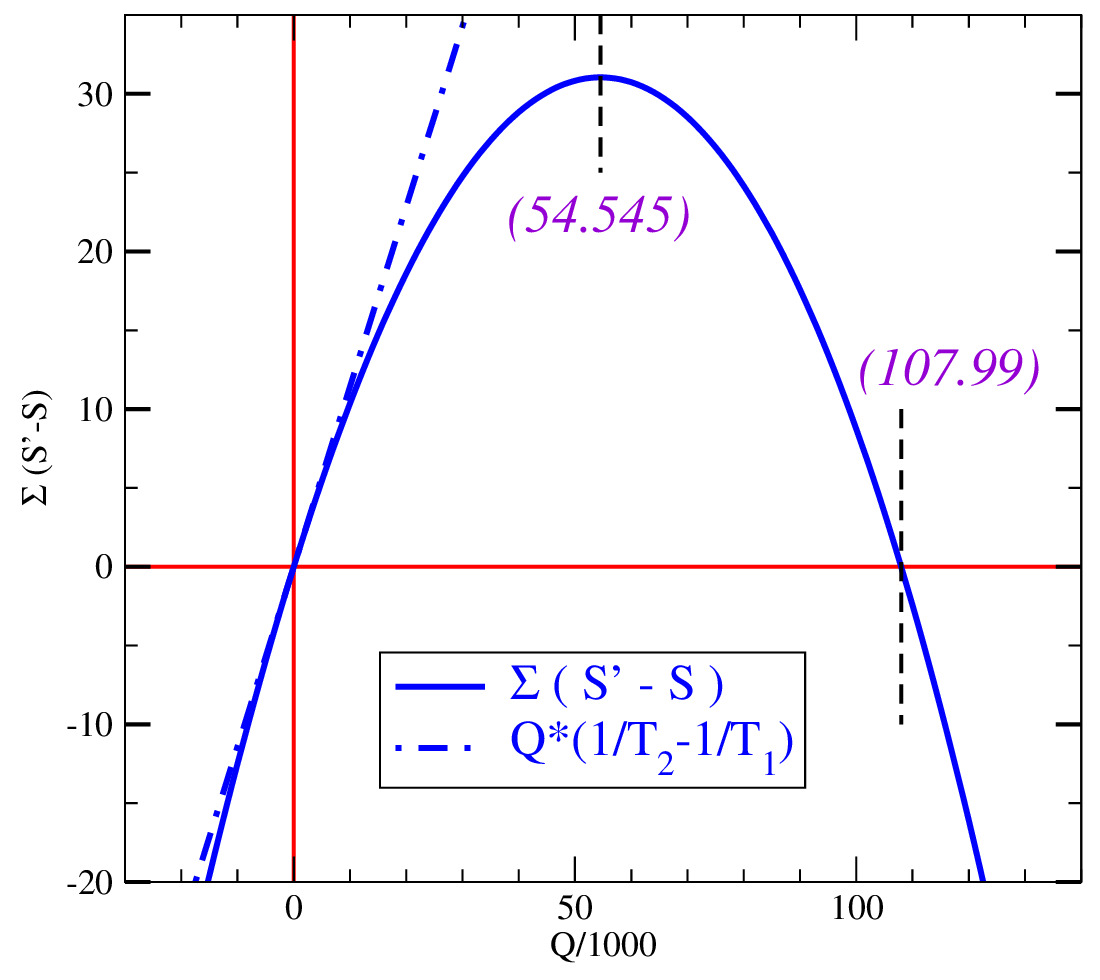} 
\vspace{-3mm}
\caption{\it\color{red} 
The graph of the function $\;\sum\phantom{s}\!\!\!(S'-S)$ of \citet{Planck_1879} given by (\ref{Eq_24}) and plotted against $Q/1000$ for the arbitrary  case $T_1=350$, $T_2=250$, $c_1=1000$ and $c_2=1200$, leading to $Q_1/1000=350$, $Q_2/1000=300$, $Q_{max}/1000 \approx 54.545$ given by (\ref{Eq_27}) for the maximum, and $Q_0/1000 \approx 107.99$ for the second numerical root (other than $Q_0=0$) of $\sum\phantom{s}\!\!\!\!=0$ given by (\ref{Eq_28}). The dashed-dotted straight line is the linear function $Q\:(1/T_2-1/T_1)$ independent on $c_2$ and $c_1$ and tangent to the curve $\;\sum\phantom{s}\!\!\!(S'-S)$ at $Q=0$. Figure plotted by P. Marquet for the present English translation of Planck's thesis.
\label{fig_Sum_Q_Planck_1879}}
\vspace*{2mm} \hrule  \vspace*{2mm}
\end{figure}

We therefore now move on to examining the function $\sum\phantom{s}\!\!\!\!$ of $Q$ given by (\ref{Eq_24}):
\begin{align}
\;\;\;\;\;\;\;\;\;\;\;\;\;\;
\;\;\;\;\;\;\;\;\;\;\;\;\;\;
\;\;\;\;\;\;\;\;\;\;\;\;
\;\;\;\;\;\;\;\;\;
 \sum \;(Q)
 & \: = \; 
  c_1\:\ln\!\left(1 \:-\: \frac{Q}{c_1\:T_1}\right) 
  \;+\;
  c_2\:\ln\!\left(1 \:+\: \frac{Q}{c_2\:T_2}\right) 
\nonumber \: .
\;\;\;\;\;\;\;\;\;\;\;\;\;\;
\;\;\;\;\; (24)
\end{align}
It is most convenient for us to imagine the corresponding values of $Q$ and $\sum\phantom{s}\!\!\!\!$ plotted as abscissas and ordinates (respectively) and look at the {\it\color{red} shape/course/trend/slope} of the curve represented in this 
way.$\,$\footnote{$\:${\it\color{red}See for instance the Fig.~\ref{fig_Sum_Q_Planck_1879} that I have plotted for the present English translation. By the way, note that Max Planck almost never plot any figures in his papers and books / P. Marquet.}}

The {\it\color{red}(following)} two differential quotients {\it\color{red}(first- and second-order derivatives)} are used for this purpose: 
\begin{align}
 \frac{d\sum\phantom{s}\!\!\!}{dQ}
 & \: = \; 
    -\:\frac{c_1}{c_1\:T_1 \: -\: Q}
  \;+\;
       \frac{c_2}{c_2\:T_2 \: +\: Q}
\label{Eq_25} \: 
\end{align}
\vspace*{-3mm}
and 
\begin{align}
 \frac{d^2\sum\phantom{s}\!\!\!}{dQ^2}
 & \: = \; 
    -\:\frac{c_1}{(c_1\:T_1 \: -\: Q)^2}
  \;-\;
       \frac{c_2}{(c_2\:T_2 \: +\: Q)^2}
\label{Eq_26} \: .
\end{align}
From the expression for $\sum\phantom{s}\!\!\!\!$ one can immediately 
see$\,$\footnote{$\:${\it\color{red} Because the arguments of the two logarithms must be positive in (\ref{Eq_24}) (P. Marquet)}.}
that the curve is real only for the values of $Q$ that lie between $-\,c_2\:T_2$ and $c_1\:T_1\:$.
According to the above a priori considerations other values of $Q$ are therefore not at all conceivable, and for each of the two limits of $Q$ we have $\sum\phantom{s}\!\!\!\!=-\:\infty$.

Furthermore, it follows from the substantially negative values of the 2nd differential quotient (\ref{Eq_26}) that the curve always has its convex side pointing upwards and therefore only has a single maximum.
For $Q=0$ the sum $\sum\phantom{s}\!\!\!\!=0$, therefore the curve intersects the abscissa axis here.
The value of $Q$ corresponding to the maximum of $\sum\phantom{s}\!\!\!\!$ results from the equation: $d\sum\phantom{s}\!\!\!\!/dQ=0$, i.e. according to (\ref{Eq_25}):  
\begin{align}
    -\:\frac{c_1}{c_1\:T_1 \: -\: Q}
  \;+\;
       \frac{c_2}{c_2\:T_2 \: +\: Q}
\; = \; 0 \: ,
\nonumber\: 
\end{align}
from which: 
\vspace*{-3mm}
\begin{align}
 Q
 & \: = \; 
  \frac{c_1\: c_2}{c_1\:+\: c_2}
  \;\left(\,T_1\:-\:T_2\,\right)
\label{Eq_27} \: ,
\end{align}
which is 
positive$\,$\footnote{$\:${\it\color{red} Because it was assumed before that $T_1>T_2$ (P. Marquet)}.}.
The curve therefore intersects the abscissa axis twice, once before the maximum, for $Q=0$, and once after the maximum.
Consequently, the equation $\sum\phantom{s}\!\!\!\!=0$, or according to (\ref{Eq_24}): 
\begin{align}
  c_1\:\ln\!\left(1 \:-\: \frac{Q}{c_1\:T_1}\right) 
  \;+\;
  c_2\:\ln\!\left(1 \:+\: \frac{Q}{c_2\:T_2}\right)
 & \: = \; 0
\label{Eq_28}
\end{align}
has, in addition to the {\it\color{red}(obvious)} root $Q=0$, a second positive root in $Q$
{\it\color{red} (See the Fig.~\ref{fig_Sum_Q_Planck_1879}).}

Now we want to examine the physical meaning of these results.
For all negative values of $Q$, from $-\,c_2\:T_2$ to $0$, $\;\sum\phantom{s}\!\!\!$ is negative, and the corresponding change of state cannot be brought about by any process, i.e. in all cases where heat is transferred from the colder to the warmer gas, no process is possible.
Nature always favours the first state more than the second.
For $Q=0$, $\;\sum\phantom{s}\!\!\!\!=0$, which corresponds to a neutral {\it\color{red}(indifferent)} process.
Then positive values of $\;\sum\phantom{s}\!\!\!$ follow, for which the corresponding changes of state can be carried out by natural {\it\color{red}(privileged)} processes.
The maximum of the entropy value-{\it\color{red} (change)} $\;\sum\phantom{s}\!\!\!$  obviously indicates the state for which nature has the greatest preference under the given conditions, i.e. which, once reached, it can no longer leave, because from it only a change of state with a negative entropy value would be conceivable.
The value of $Q$ corresponding to this maximum value-{\it\color{red} (change)} is given in (\ref{Eq_27}), and if this value-{\it\color{red} (change)} is used to calculate the final temperatures $T'_1$ and $T'_2$ corresponding to it from (\ref{Eq_22}) and (\ref{Eq_23}), we find: 
\vspace*{0mm} 
\begin{align}
 T'_1 & \: = \; 
 \frac{c_1\:T_1 \:+\: c_2\:T_2}{c_1\:+\: c_2}
 \nonumber \: ,
\\
 T'_2 & \: = \; 
 \frac{c_1\:T_1 \:+\: c_2\:T_2}{c_1\:+\: c_2}
 \nonumber \: ,
\end{align}
and thus $T'_1=T'_2$. 
In fact, according to experience, this is the absolute state of equilibrium towards which nature is striving.

If $Q$ grows even more, whereby $T'_2>T'_1$, then $\;\sum\phantom{s}\!\!\!$ initially remains positive, but decreases continuously, and thus the corresponding changes of state can still be caused by natural {\it\color{red}(privileged)} processes, which would also be easy to prove in reality.
However, this is then no longer possible through direct heat exchange between the two gases, because this path would only lead through the state corresponding to the \dashuline{maximum entropy}-{\it\color{red} (change)}, from which no further change is possible.
In order to set up such a natural {\it\color{red}(privileged)} process, one would either have to temporarily change the volume of one of the two gases, or one could also leave the volumes constant and draw a third gas into the process, which would have to be exactly in its original state again at the end of the process.

The curve then intersects the abscissa axis for the second time
{\it\color{red} (see for example the Fig.~\ref{fig_Sum_Q_Planck_1879})}.
This is the boundary {\it\color{red}(limit)} point discussed above, at which nature again has an equal preference for both states.
The corresponding value of $Q$ is the second  
root$\,$\footnote{\it\color{red}$\:$As an example, see the second non-trivial root value $Q_0/1000 \approx 107.99$ in the Fig.~\ref{fig_Sum_Q_Planck_1879}  computed iteratively from (\ref{Eq_28}) via a Newton's method  (P. Marquet).}
of the transcendental equation $\;\sum\phantom{s}\!\!\!\!=0$  given by (\ref{Eq_28}).
It would be easy to show that for this value of $Q$ the two states can really be transformed into each other by a neutral {\it\color{red}(indifferent)} process by organising a reversible process, which then requires the condition given in (\ref{Eq_28}).
However, the condition given in (\ref{Eq_28}) would require us to do so.
For the special case $c_1=c_2$ the equation (\ref{Eq_28}), apart from the root $Q=0$, gives the value: 
$$Q\;=\;c_1\:(\,T_1\:-\:T_2\,) \: ,$$ 
from which by means of (\ref{Eq_22}) and (\ref{Eq_23}) the final temperatures result: 
$$T'_1\;=\;T_2 \;\;\;\mbox{and}\;\;\; T'_2\;=\;T_1 \: ,$$ 
i.e. the temperatures have just exchanged as predicted above.

For even larger values of $Q$, $\;\sum\phantom{s}\!\!\!$ becomes negative again and decreases continuously {\it\color{red} (see for example the right part of the Fig.~\ref{fig_Sum_Q_Planck_1879})} until it reaches the value $-\:\infty$ for $Q=c_1\:T_1$.
For all these cases, no process in the direction from the first to the second state is possible.

It is remarkable that for very small values of $Q$ the entropy value-{\it\color{red} (change)} $\;\sum\phantom{s}\!\!\!$ only depends on $Q$ and the temperatures $T_1$ and $T_2$, but not on the heat capacities $c_1$ and $c_2$.
Indeed, if we assume $Q$ to be infinitely small, the equation (\ref{Eq_24}): 
\begin{align}
 \sum 
 & \: = \; 
  c_1\:\ln\!\left(1 \:-\: \frac{Q}{c_1\:T_1}\right) 
  \;+\;
  c_2\:\ln\!\left(1 \:+\: \frac{Q}{c_2\:T_2}\right) 
\nonumber \: 
\end{align}
is transformed into : 
\vspace*{-3mm}
\begin{align}
 \sum 
 & \: = \; 
  c_1\:\left(\:-\: \frac{Q}{c_1\:T_1}\right) 
  \;+\;
  c_2\:\left(\:+\: \frac{Q}{c_2\:T_2}\right) 
\nonumber \: 
\end{align}
or: 
\vspace*{-3mm}
\begin{align}
 \sum 
 & \: = \; Q \;
  \left( \frac{1}{T_2} \:-\: \frac{1}{T_1}\right) 
\nonumber \: 
\end{align}
{\it\color{red} (see for example the Fig.~\ref{fig_Sum_Q_Planck_1879} and the dashed-dotted straight line tangent to the curve $\;\sum\phantom{s}\!\!\!(Q)$ at the origin $Q=0$ / P. Marquet).}

Now, let us determine the maximum of the entropy value-{\it\color{red} (change)}.

\vspace*{-1mm} 
\begin{center}
-------- \underline{Second Example (p.26-27)} -------- 
\end{center}
\vspace*{-2mm}

We again assume two perfect gases in the same state as in the previous example, but this time at the same temperature $T$, and want to calculate the entropy value-{\it\color{red} (change)} due to the change in state that occurs as a result of the change of the volumes $v_1$ and $v_2$, while the temperature of both gases in the second state should be the same as in the first.
If the volumes in the second state are $v'_1$ and $v'_2$, the entropy value-{\it\color{red} (change)} of the transition from the first to the second state is obviously:
\begin{align}
 \sum \left( S' \, - \, S \right) 
 & \: = \; 
  k_1\:\ln\!\left(\frac{v'_1}{v_1}\right) 
  \;+\;
  k_2\:\ln\!\left(\frac{v'_2}{v_2}\right) 
 \: .
\label{Eq_29}
\end{align}
If $v'_1>v_1$ and $v'_2>v_2$, then $\;\sum\phantom{s}\!\!\!$ is positive, i.e. a natural  {\it\color{red}(privileged)}  process is possible, as can be seen immediately from the fact that both gases are allowed to expand without {\it\color{red} overcoming} pressure.

However, if for example $v'_1>v_1$ and $v'_2<v_2$, then $\;\sum\phantom{s}\!\!\!$ can be positive or $=0$.
In the latter case, a neutral {\it\color{red}(indifferent)}  process between the two states is possible, and the expansion of the first gas can then be regarded as the compensation required to reduce the volume of the second.

In the following, however, we will limit ourselves to determining the state that is most favoured by nature under the given conditions, i.e. from which no natural {\it\color{red}(privileged)} process is possible.
This state corresponds to the \dashuline{\,maximum of the entropy value} and it is a matter of finding the corresponding values of $v'_1$ and $v'_2$.
If these two quantities are left completely arbitrary, it follows that $\;\sum\phantom{s}\!\!\!$ can assume infinitely large values, since in fact both gases can continually increase their volume by natural {\it\color{red}(privileged)} processes without {\it\color{red} overcoming} pressure.
We therefore want to set the condition that the total volume of both gases is constant, i.e.: 
$$ v'_1 \;+\; v'_2 \;=\; v_1 \;+\; v_2 \: ,$$
from which, with $d(v_1+v_2)=0$:
\vspace*{-3mm}
\begin{align}
 dv'_1 \;+\; dv'_2 & \: = \; 0 \: .
\label{Eq_30}
\end{align}

Let us now determine the maximum of the entropy-{\it\color{red}(change)} value $\sum\phantom{s}\!\!\!$.
For this, according to (\ref{Eq_29}) and since $v_1$ and $c_2$ are constant, 
\begin{align}
 d \:\sum\phantom{s}\!\!\!
 & \: = \; 
  k_1\;\frac{dv'_1}{v'_1}
  \;+\;
  k_2\;\frac{dv'_2}{v'_2}
  \; = \; 0
 \: 
\nonumber
\end{align}
follows, and taking into account (\ref{Eq_30}): 
\vspace*{-3mm}
\begin{align}
 \frac{k_1}{v'_1} 
 & \: = \;  
 \frac{k_2}{v'_2} 
 \; ,
\label{Eq_31}
\end{align}
and thus also 
\vspace*{-3mm}
\begin{align}
 k_1\;\frac{T}{v'_1} 
 & \: = \;  
 k_2\;\frac{T}{v'_2} 
 \; ,
\nonumber
\end{align}
or
\begin{align}
 p'_1
 & \: = \;  
 p'_2
 \; ,
\nonumber
\end{align}
i.e. the pressures of the two gases tend to equalise, as experience also teaches, if the two gases are brought into such contact that their pressure forces can act on each other, while keeping their total volume constant.

\vspace*{-1mm} 
\begin{center}
-------- \underline{Third Example (p.27-28)} -------- 
\end{center}
\vspace*{-2mm}

We want to consider a very simple case in which mechanical work is also carried out.
If for this purpose we imagine a single perfect gas in $2$ different states, which are to be characterised by the volumes $v$, $v'$ and by the temperatures $T$, $T'$, and if, for the sake of comparability of these two states, we have to imagine a corresponding mechanical workload-supply included in one of them, the entropy-{\it(\color{red}change)} value of the transition from the first to the second state is:
$$ S' \;-\; S \;=\; 
c\:\ln\left(\frac{T'}{T}\right)
\;+\;
k\:\ln\left(\frac{v'}{v}\right) \; .$$
It follows from this that the transition can be accomplished by a natural {\it\color{red}(privileged)} process if $T'>T$ and $v'>v$ {\it\color{red}(implying $S'-S>0$ if the two conditions are fulfilled)}.
In fact, if we allow the gas to expand as desired without overcoming any pressure, we obtain a larger volume at the same temperature.
If we then return the gas to its old volume by compression, we have an increased temperature at the same volume.
But if you want to reduce the temperature of the gas so that $T'<T$, then the volume must necessarily increase accordingly as compensation, and vice versa. 
However, it will never be possible to lower the temperature of the gas while maintaining the same volume, or to reduce the volume at the same temperature, without causing changes in other bodies, because such a change in state has a negative entropy-{\it\color{red}(change)} value and cannot therefore be carried out without another natural {\it\color{red}(privileged)} process that serves as compensation.

\subsection{\underline{Arbitrary bodies} (p.28-43)}
\vspace*{-1mm}

The generalization of the developed-{\it\color{red}(previous)} theorem to any field requires first proving the existence of the entropy function for every field, i.e. the proof that for every body the expression 
\begin{align}
 \frac{\;dU \;+\: p \: . \: dv\;}{T} 
 \; ,
\label{Eq_32}
\end{align}
where $dU$ means the increase in energy (heat + internal work) and $p$ means the pressure of the body towards the outside (both thought of as functions of $T$ and $v$), is the complete differential {\it\color{red}(namely $dS$)} of a function of $T$ and $v$: the Entropy.
Its value {\it\color{red}(of the entropy)} is completely determined by the current state of the body.
\dashuline{\,Clausius} provided this proof in the first form in which he published the \dashuline{\,second law} of the mechanical theory of heat by considering a \dashuline{\:circular-{\it\color{red}(cyclic)} process\:} and examining the various types of transformations that occur in it.
We want to use the same circular-{\it\color{red}(cyclic)} process for the same proof, but \dashuline{\,base the proof on significantly different tools}, namely on the results previously obtained when considering perfect gases, as the theorems developed there find an interesting application in this proof.

\vspace*{-1mm} 
\begin{center}
\underline{Proof of the second law in its application to a circular-{\it\color{red} cyclic} process}.
\end{center}
\vspace*{-2mm}

We imagine any (reversible or non-reversible) circular-{\it\color{red}(cyclic)} process being carried out with any body, namely in such a way that the body is subjected to any pressure forces, compressed and dilated, at the same time brought into contact with various heat reservoirs, some warmer, some colder, and finally returned to its initial state, i.e. to its original volume and temperature, by all these changes.

For the sake of generality, let us assume at once that each heat reservoir is in communication with the body for an infinitesimally short time only, which obliges 
us to suppose an infinite number of heat reservoirs.
Let us further assume that all heat reservoirs consist of containers of perfect gases, the volumes of which are kept constant. 
This restriction will not detract from the necessary scope of the proof.

At the beginning of the cycle, let the amount of heat contained in any reservoir be $Q$, its temperature be $T$. During the process, the reservoir releases the amount of heat $dQ$ to the body, thereby reducing the temperature to $T-dT$.
So $dQ$ and $dT$ are either both positive or both negative. 
If the true {\it\color{red}(specific)} heat capacity of the gas (based on its mass) is $c$, then one has: 
\begin{align}
 dQ \;=\: c \: . \: dT \; .
\label{Eq_33}
\end{align}

The cycle forms a natural {\it\color{red}(privileged)} or a neutral {\it\color{red}(indifferent)} process. Let us compare the initial state and the final state of the same with each other.
Since the body undergoing the various changes is in exactly the same conditions in both states, it can be completely disregarded when comparing the two states.
The only change that has occurred is that: 1) the heat reservoirs have changed state; and 2) some mechanical work has been done.

Both states therefore comprise only perfect gases and a mechanical work supply, and since the final state has emerged from the initial state by a process, we can apply the entropy theorem to this process, which states that this entropy value is always positive, or $=0$ in the case of a neutral {\it\color{red}(indifferent)} process, whereby the mechanical work performed does not enter into the entropy value at all.

If we therefore form the entropy value-{\it\color{red}(change)}  $\sum \left( S' \, - \, S \right)$, we first have for any single heat reservoir, since its initial temperature is $T$ and its final temperature is $T-dT$, and while its volume remains constant: 
\begin{align}
S' \: - \: S 
& \:=\; c \; \ln\!\left(\frac{T-dT}{T}\right)
  \;=\; c \; \ln\!\left(1 - \frac{dT}{T}\right)
  \;\approx\; - \: c \; \frac{dT}{T}
  \;=\; - \: \frac{c \:.\: dT}{T}
\nonumber \: ,
\end{align}
but according to (33) 
\vspace*{-3mm}
\begin{align}
  c \: . \: dT  \;=\: dQ  \; ,
\nonumber
\end{align}
and therefore for a single heat reservoir: 
\vspace*{-3mm}
$$ S' \: - \: S \;=\; - \:\frac{dQ}{T} \: . $$
Thus the entropy value-{\it\color{red}(change)} of the cyclic process is 
\begin{align}
 \sum \left( S' \, - \, S \right) 
  & \:=\: - \,\int \frac{dQ}{T}  \; > \; 0 \; ,
\label{Eq_34}
\\
 \mbox{or}\;\;\;
  \int \frac{dQ}{T}  \; < \; 0 \; ,
\label{Eq_35}
\end{align}
where the integral is to be extended over all reservoirs.

This equation is the one in which Clausius first published the second law, {\it\color{red}where} $dQ$ denotes the amount of heat absorbed by the body at any given time, $T$ its temperature, and the integral is to be extended over all amounts of heat absorbed (and released).

If we now assume the special case that the circular-{\it\color{red}(cyclic)} process is reversible, i.e. that:
 1) the external pressure forces acting are always equal to those of the body; and 
 2) each heat reservoir has the same temperature as the body at the moment of its effectiveness, we obtain a neutral {\it\color{red}(indifferent)} process and thus the equation:
\vspace*{-3mm}
\begin{align}
  \int \frac{dQ}{T}  \; = \; 0 \; .
\label{Eq_36}
\end{align}
In this case, however, $T$ also means the temperature of the body at the time of absorption of the quantity of heat $dQ$, and the latter then has the value 
$$\;\;\;\;
     dQ \;=\; dU \;+\; p \: . \: dv  \; , 
  \;\;\;\;\;\;\;\;\;\;\;\;
     \mbox{cf. (32)}$$  
so that the resulting equation 
$$ \int \frac{dU \;+\; p \: . \: dv}{T} \;=\; 0 $$ 
is composed only of characteristics of the body (functions of the independent variables $T$ and $v$).
The upper limit of the integral is identical to the lower limit because the initial and final states of the body are the 
same.$\,$\footnote{$\:${\it\color{red}Therefore that, Planck thought about the loop integral  
$\oint [\:{dU \;+\; p \: . \: dv}\:]\,/\,{T} \;=\; 0$.}}
It immediately follows that if one integrates from the initial state to any other state, the value of the integral is a very specific one, depending only on the limits, or in other words that the expression $(dU+p\:dv) /T$ is a complete differential, so that we can write 
\vspace*{0mm}
\begin{align}
  \frac{dU \:+\: p\:dv}{T} \; = \; dS \; .
\label{Eq_37}
\end{align}
The \dashuline{\,entropy $S$ is therefore completely determined} (\dashuline{\,except for an additive constant}) by the state of the body, and this is the result we had to draw from the above 
proof.$\,$\footnote{$\:${\it\color{red}I remember in the Preface the words of Max Planck, who wrote in 1943 about his thesis of 1879 and the concept of entropy, that: ``\dashuline{\,Kirchhoff explicitly rejected its content\,}, stating that \dashuline{\,the concept of entropy, whose magnitude can only be measured and therefore defined by a reversible process, should not be applied to irreversible processes\,}''. 
This just shows that anyone can make a mistake, even the best of us\,! And this shows that the entropy function is definitely very difficult to understand, as is the fact that the third law (entropy of the more stable solids is zero at $0$~K) must be applied as soon as we study bodies of varying composition, such as the atmosphere (more or less humid) or the ocean (more or less salty).}}

We are now in a position to repeat all the observations we made with regard to the entropy-{\it\color{red}(change)} value of a process with perfect gases, in a very similar way but here with arbitrary bodies, and thus to repeat and generalise the results found there {\it\color{red}(i.e. for perfect gases)}.

Since the treatment of the individual cases is completely analogous to that there {\it\color{red}(for perfect gases)}, we can confine ourselves here to briefly stating the main points of view and otherwise referring to the explanations there {\it\color{red}(made for perfect gases)}.

Firstly, we return to the comparison of states that are infinitesimally different from one another.

\subsubsection{\underline{Comparison of two states of a single body} (p.32-33)}
\vspace*{-2mm}

The two given states are: 
\vspace*{1mm} \\ \hspace*{26mm}
1st State): 
temperature $\;\;\;\;\;T$, 
\;\;\;\;volume $\;\;\;\;v$ ; 
\vspace*{1mm} \\ \hspace*{25mm}
2nd State): 
temperature $T+dT$, 
volume $v+dv\:$.

To ensure that the two states contain the same supply of force, we add to the first an amount of mechanical energy of the magnitude $dU$ (the difference between the energies of the body in the two states).

The sufficient condition for the possibility of a neutral {\it\color{red}(indifferent)} process results from the execution of a reversible process from the first state, which should bring the body to the volume $v+dv$, whereby it may reach the temperature $T+\Delta T$.
Then we have, since the heat added from outside is $0$:
\vspace*{0mm} \begin{align}
  0 \; = \; \Delta U \:+\: p \:.\: dv \; ,
\label{Eq_38} \end{align}
where $\Delta U$ is the increase in energy caused by the changes $\Delta T$ and $dv$ in temperature and volume, and where $p$ is the pressure of the body towards the outside.
A transition from this state $T+\Delta T$, $v+dv$ to the second given state is in any case possible by a neutral {\it\color{red}(indifferent)} process if $\Delta T = dT$.
Thus we obtain by substitution of $dT$ instead of $\Delta T$ in (\ref{Eq_38}), as a \dashuline{\,sufficient condition of a neutral {\it\color{red}(indifferent)} process}: $dU + p\:dv=0$, or by division with $T$ and according to (\ref{Eq_37}): 
\vspace*{-1mm} \begin{align}
  \boxed{\: dS \; = \; 0 \;} \: .
\label{Eq_39} \end{align}

Let us now turn to the establishment of the sufficient condition of a natural {\it\color{red}(privileged)} process from the first to the second state.
As with perfect gases, everything that follows is based on the assumption that a natural {\it\color{red}(privileged)} process occurs when a body expands without overcoming external pressure, i.e. with constant energy  (see the note~\ref{note_p10}).
So if $dU=0$ and for a $dv$ given positive, we would have a natural {\it\color{red}(privileged)} process from the first to the second state.
Then, since according to (\ref{Eq_37}) 
$dS=(dU+p\:dv)/T$, for this case 
$dS=p\:dv/T$ would be essentially positive, 
leading to: 
\vspace*{-1mm} \begin{align}
  \boxed{\: dS \; > \; 0 \;} \: .
\label{Eq_40} \end{align}
From this it can be proven that \dashuline{\,this is also in general the sufficient condition for the possibility of a natural {\it\color{red}(privileged)} process from the first to the second state},
because if the two given states fulfil this condition, the entropy of the body can be brought from the first state to the larger value $S+dS$ by a natural {\it\color{red}(privileged)} process of the kind just considered (by expansion of the body without overcoming pressure), and then the second state can be brought about by a neutral {\it\color{red}(indifferent)} process while maintaining this value of entropy.
Then the whole transition took place through a natural {\it\color{red}(privileged)} process.

Likewise, $\boxed{\:dS<0\:}$ is the \dashuline{\,sufficient condition for the possibility of a natural {\it\color{red}(privileged)} process from the second to the first state} (i.e. for the impossibility of a process from the first to the second state).

Finally (\ref{Eq_39}) and (\ref{Eq_40}) also emerge as the \dashuline{\,necessary conditions for a neutral {\it\color{red}(indifferent)} and a natural {\it\color{red}(privileged)} process from the first to the second state}.

\subsubsection{\underline{Comparison of two states of two bodies} (p.33-36)}
\vspace*{-2mm}

The two given states are: 
\vspace*{1mm} \\ \hspace*{26mm}
1st State): 
\;\;\;\;\;\; 
$T_1$ \;$v_1$ \;$S_1$, 
\;\;\;\;\;\; 
$T_2$ \;$v_2$ \;$S_2$
\vspace*{1mm} \\ \hspace*{25mm}
2nd State): 
\;\;\;\;\;\;\;\;$S_1+dS_1$, 
\;\;\;\;\;\;\;\;\;$S_2+dS_2$
\vspace*{1mm} \\ 
whereby the first state still has to be provided with the corresponding mechanical work supply $(=dU_1+dU_2)$.

The procedure can also be applied to perfect gases by first bringing both bodies individually to the same temperature $T$ through reversible compression to determine the sufficient condition for the possibility of a neutral {\it\color{red}(indifferent)} process.
However, one could see a difficulty in carrying out this process, especially if the given temperatures $T_1$ and $T_2$ are very far apart.
We would therefore like to adopt another method of proof, which requires only minimal changes in the temperatures of both bodies and which could also have been used above for the perfect gases.
It is also based on the execution of a neutral {\it\color{red}(indifferent)} process that, starting from the first state, should bring the two bodies to the entropies $S_1+dS_1$ and $S_2+dS_2$.

We use an arbitrary quantum {\it\color{red}(\,``{\it\,Quantum\,}'' in German)} of a perfect gas that must mediate/help {\it\color{red}(\,``{\it\,Hilfe\,}'' in German)} the process, starting from the initial temperature $T_1$ of the first 
body.$\,$\footnote{$\:$\it\color{red}Note that it has often been suggested that Planck has chosen in 1900 the letter ``\,$h$\,'' for denoting his fundamental new universal constant for indicating a  desperate action to help (``{\it\,\underline{h}ilfe\,}'') the introduction of ``{\it\,Quantum\,}'' made of quantized  and discrete values of exchange of energy (P. Marquet).}
We bring this gas into a thermally conductive connection with the first body and allow it to expand or contract in this connection in a reversible manner.
This changes the entropy of the first body and it can therefore be brought to the value $S_1+dS_1$.
The mediating gas is then isolated and brought to the temperature $T_2$ of the second body through a reversible process without the addition of heat from outside.
Next it is placed in a heat-conducting connection with the latter, and thus brought to the entropy $S_2+dS_2$ in a reversible manner.
Finally the gas is possessed again, so it can be brought back to its original state through a reversible process, and then the final state of the entire reversible process is the required one  (because the mediating gas behaves exactly as it did at the beginning and the two given bodies have been converted into the required states).
But if this condition is not fulfilled, then exactly the second given state has not been brought about, because a third body has suffered changes in addition to the given ones.
We now want to set up the corresponding equations.

In the first goal/target of the reversible process, the gas is in a thermally conductive connection with the first body, both at the temperature $T_1$.
Let the entropy $S$ of the gas change to $S+dS$. Then, when the amount of heat $dQ$ has passed from the body into the gas: 
\vspace*{1mm} \\ \hspace*{50mm}
for the gas  \;\;\;\;  $dQ = T_1 \: dS \; ,$ 
\vspace*{1mm} \\ \hspace*{46mm}
for the body \;\;  $-\:dQ = T_1 \: dS_1 \; ,$ 
\vspace*{1mm} \\ 
and from this:
\vspace*{-3mm} \begin{align}
  dS \:+\: dS_1 \; = \; 0 \; .
\label{Eq_41} \end{align} 
As a result, while the gas is brought to temperature $T_2$, its entropy $S+dS$ remains unchanged.
But then it is placed in a heat-conducting connection with the second body, whereby it assumes the entropy $S+dS+\Delta S$.
Then we have analogous to equation (\ref{Eq_41}): 
\vspace*{0mm} \begin{align}
  \Delta S \:+\: dS_2 \; = \; 0 \; .
\label{Eq_42} \end{align}
Now if the gas has its initial entropy $S$ again, i.e. if: 
\vspace*{0mm} \begin{align}
  d S \:+\: \Delta S \; = \; 0 \; ,
\label{Eq_43} \end{align}
then, by adding (\ref{Eq_41}) and (\ref{Eq_42}): 
\vspace*{-5mm} \begin{align}
  \boxed{\:dS_1 \:+\: dS_2 \; = \; 0 \:} \; 
\label{Eq_44} \end{align}
is, according to the above explanations, the \dashuline{\,sufficient condition for the possibility of a neutral process {\it\color{red}(indifferent)} from the first to the second state}.
But if equation (\ref{Eq_43}) is not fulfilled, and if for instance at the end of the reversible process the entropy of the gas is smaller than at the beginning, i.e. if 
\vspace*{0mm} \begin{align}
  d S \:+\: \Delta S \; < \; 0 \; ,
\label{Eq_45} \end{align}
then you can obviously bring about the second given state by finally letting the gas carry out a natural {\it\color{red}(privileged)}  process for itself, and then its own Entropy increases again up to $S$, and thereby returns it to its original state.
Then the process carried out as a whole was a natural  {\it\color{red}(privileged)}  one. Therefore, the prerequisite (\ref{Eq_45}), which transforms the equation (\ref{Eq_44}) into
\vspace*{0mm} \begin{align}
  \boxed{\: d S_1 \:+\:d S_2 \; > \; 0 \:} \; ,
\label{Eq_46} \end{align}
provides us with the \dashuline{\,sufficient condition for the possibility of a natural {\it\color{red}(privileged)} process from the first to the second state}.
The following conclusions are the same as above for perfect gases.

\subsubsection{\underline{Comparison of two states of several  bodies} (p.36-43)}
\vspace*{-2mm}

The remaining essential discussions, namely the generalization to several bodies and to finite differences in states, can be transferred here directly from above, so that we can content ourselves with immediately stating the general result on which our entire argument culminates.

If of any number of arbitrary bodies any two different states are given, which, however, comprise the same stock of matter and force, and the entropy of any one of these bodies is $S$ in the first state and $S'$ in the second state, we call the expression 
\vspace*{0mm}  
\begin{align}
 \sum \left( S' \, - \, S \right) 
\label{Eq_47}
\end{align}
the entropy-{\it\color{red}(change)} value of the transition from the first to the second state.
Here the sum symbol is to be extended to all bodies which are in different conditions in both states. On the other hand, mechanical reserves of work and mechanical living force {\it\color{red} (kinetic energy)}  are not included in this expression.
Then \dashuline{\,we have the following {\it\color{red}Entropy-}{\bf theorem}:\,} 
\vspace*{-2mm}
\begin{quotation}
{\bf Depending on whether the \dashuline{\,entropy-{\it\color{red}(change)} value\,} $=0$ or is positive or negative, the transition between the two states can be accomplished in both directions, or only in the direction from the second to the first state in nature.
The first case corresponds to neutral {\it\color{red}(indifferent)} processes, the last two to natural {\it\color{red}(privileged)} processes.}
\vspace*{2mm}

{\bf Conversely: the \dashuline{\,entropy-{\it\color{red}(change)} value\,} of every process is positive or $=0$.
In the first case the process is natural {\it\color{red}(privileged)}, in the second it is neutral {\it\color{red}(indifferent)}.}
\end{quotation}
\dashuline{\,This theorem is to be viewed as the most general form of the second law\,} of the mechanical theory of heat.
The reversal was already expressed by Clausius in the sentence: ``\,The entropy of the world strives/tends towards a maximum,'' \dashuline{\,while the first form, which implies an even greater generality, has not yet been established as far as I know\,}.
Therefore, in order to bring about a change of state with negative entropy-{\it\color{red}(change)} with one group of bodies, another related change of state with positive entropy-{\it\color{red}(change)} is required with another group of bodies, in such a way that the entropy-{\it\color{red}(change)} of the total change is positive or at least $=0$.
One can therefore consider the absolute value of the negative entropy-{\it\color{red}(change)} value of the change of state to be carried out as the measure of the compensation necessary to make it possible, and the second discussed change of state as a compensation.

In general, however, it must be noted that if there is a change of state with a positive entropy-{\it\color{red}(change)} value, which can therefore take place in nature, not every arbitrary type of transition is actually carried out by nature, but that, for example, equilibrium can very well exist even if the absolute maximum of the entropy-{\it\color{red}(change)} value has not yet been reached
(because the \dashuline{\,entropy theorem} only states that such a change of state can always be brought about by a suitable arrangement of the external conditions of the system of bodies).
We first want to apply the above statement to two simple cases, namely heat conduction and friction, in order to check its agreement with direct experience.

\vspace*{1mm} 
\begin{center}
-------- \underline{First example: Heat conduction (p.38-41)} -------- 
\end{center}
\vspace*{-2mm}

Let us imagine any two different bodies whose states are given.
The aim is to determine the entropy value of the change in state that occurs when a quantity of heat $Q$ transfers from the 1st body to the 2nd.

This task is indefinite in its given form, because the entropy value: 
$$(\,S'_1\:-\:S_1\,) \;+\; (\,S'_2\:-\:S_2\,)$$ 
(where the indices refer to the two bodies) requires complete knowledge of the final state in order to be determined, and this is obviously not determined by the transferred heat $Q$ alone, but also through the other possible changes in the state of the two bodies, such as: changes in volume or pressure, etc.

In order to fix the problem a little more we add the condition that, in the event of any changes in volume of the two bodies, the external pressure must always be set equal to the counter-pressure of the body in question, and in other words: that the volume changes take place in a reversible manner.
Even now, the problem is generally still indeterminate, depending on whether the heat transfer takes place at constant volume or at constant pressure, for example. However, in this case, for an infinitely small value of $Q$, it can be shown that the problem has a specific solution.
If we allow the infinitely small quantity of heat $dQ$ to pass from the first to the second body, whose initial temperatures may be $T_1$ and $T_2$, we have, owing to the assumed condition that external pressure and counter-pressure are equal: 
\vspace*{-1mm}  
\begin{align}
\mbox{For the first body:} \;\;\; -dQ & \: = \; T_1 \:.\: dS_1
\label{Eq_48} \: , \\
\mbox{For the second body:} \;\;\; +dQ & \: = \; T_2 \:.\: dS_2
\label{Eq_49} \: .
\end{align}
Thus the entropy value of the change of state
\vspace*{-1mm}  
\begin{align}
dS_1 \:+\: dS_2 & \: = \; 
dQ \:.\: \left( 
\:\frac{1}{T_2} \:-\: \frac{1}{T_1}\:
\right)
\label{Eq_50} \:
\end{align}
is completely independent of the volumes of both bodies, as well as of the temperature changes that occur. 
If $dQ$ is positive, then this expression is positive for $T_1>T_2$ ; i.e. the heat transfer from a warmer to a colder body forms a natural {\it\color{red}(privileged)} process.
However, the above entropy-{\it\color{red}(change)} value (\ref{Eq_50}) is obviously only valid if 1) in the case of volume changes the external pressure is to be set equal to the counter-pressure of a body, and if 2) the amount of heat transferred is infinitely small (because for finite state changes the entropy value is also dependent on the simultaneous volume changes of the two bodies).
In fact, in order to integrate equation (\ref{Eq_50}), it is necessary to know the dependence of the temperature on $Q$, and therefore also the changes in volume.

We now want to analyse the conditions under which the expression $(S_1+S_2)$ reaches a maximum.
This maximum will then give us the state from which no process is possible under the given conditions of the task, and in which the two bodies must persist once they have reached it (absolute equilibrium state).
Let us again assume that changes in the volume of bodies can only ever occur in a reversible manner.

Then we can use the equation 
\vspace*{-3mm}  
\begin{align}
T_1 \;\, dS_1 \;+\; T_2 \;\, dS_2 & \: = \; 0
\label{Eq_51} \:
\end{align}
which is obtained by adding (\ref{Eq_48}) and (\ref{Eq_49}).

A necessary but generally not sufficient condition of the maximum of $S_1+S_2$ is that 
\vspace*{-1mm}  
\begin{align}
 dS_1 \;+\; dS_2 & \: = \; 0
\label{Eq_52} \:
\end{align}
for all possible infinitely small changes in the variables that determine the state of the two bodies.
Equations (\ref{Eq_51}) and (\ref{Eq_52}) combined result in 
\vspace*{-1mm}  
\begin{align}
 T_1 \; = \; T_2
\label{Eq_53} \:
\end{align}
as a necessary condition of the maximum, i.e. the absolute equilibrium state, and this is consistent with direct experience.
However, one can also see from direct observation that this condition is not yet sufficient, since from such a state in which the bodies have the same temperatures, the conditions assumed are still compatible.
One only needs to think of both bodies isolated and each of them reversibly compressed or expanded in such a way that their temperatures become different. 
If we then allow the bodies to exchange heat again, we have a natural {\it\color{red}(privileged)} process according to (\ref{Eq_50}) and therefore an increase in the entropy sum $(S_1+S_2)$ ; in other words, the maximum has not yet been reached.

A sufficient condition of the maximum is that, in addition to the equation (\ref{Eq_52}) $dS_1+dS_2=0$, the following inequality applies 
\vspace*{-3mm}  
\begin{align}
d\,^2S_1 \;+\; d\,^2S_2  \; < \; 0
\label{Eq_54} \:
\end{align}
for all possible changes in the variables.
However, the value of the latter expression can easily be found by differentiating equation (\ref{Eq_51}) 
$T_1 \; dS_1 \,+\, T_2 \; dS_2 \: = \: 0$, 
which is always valid.
We then obtain 
$$  T_1\:d\,^2S_1 \;+\; T_2\:d\,^2S_2  
 \;+\; dT_1\;dS_1 \;+\; dT_2\;dS_2 \;=\; 0$$ 
and since, according to (\ref{Eq_53}), $T_1=T_2$: 
$$ d\,^2S_1 \;+\; d\,^2S_2  \;=\; 
-\:\frac{dT_1\;dS_1 \;+\; dT_2\;dS_2}{T_1}\;.$$
If you now set $dS_1=-\,dQ/T_1$ after (\ref{Eq_48}) and $dS_2=+\,dQ/T_2$ after (\ref{Eq_49}), it results 
$$ d\,^2S_1 \;+\; d\,^2S_2  \;=\; -\:\frac{dQ}{(T_1)^2} 
   \; \left( dT_2 \;-\; dT_1 \right)\: ,$$ 
and since the left-hand side of the equation is negative, then 
$$dQ \: . \: \left( dT_2 \:-\: dT_1 \right)$$
must be positive.

We therefore have a \dashuline{\,sufficient condition for the maximum of $(S_1+S_2)$}, i.e. the  \dashuline{\,absolute state of equilibrium}: 
\vspace*{1mm} \\
\hspace*{20mm} 1) \hspace*{20mm} 
$T_1 \; = \; T_2$ \: ,
\vspace*{1mm} \\
\hspace*{20mm} 2) \hspace*{20mm} 
$dQ \: . \: \left( dT_2 \:-\: dT_1 \right)>0$ \, .
\vspace*{2mm} \\
In words: The two bodies in question have reached any case of absolute equilibrium if 1) their temperatures are the same, and 2) any conceivable change of state would cause a heat transfer in such a way that the body receiving the heat would become warmer if the body receiving the heat became warmer than the other 
(because if $dQ$ is positive, then $dT_2-dT_1$ must also be positive and vice versa).
If the 2nd condition is not fulfilled, which is the case, for example, if a change of state is allowed in which $dQ=0$ or $dT_2-dT_1=0$, then the maximum and therefore also the absolute state of equilibrium need not occur, as can also be seen from direct observation. 


\vspace*{-1mm} 
\begin{center}
-------- \underline{Second example: Friction and impact/shock (p.41-42)} -------- 
\end{center}
\vspace*{-2mm}

When two bodies rub or push against each other, heat is generated at the expense of their living force {\it\color{red} (kinetic energy)}.
We want to calculate the entropy-{\it\color{red}(change)} value of such a process by assuming that the lost living force {\it\color{red} (kinetic energy)} has only been converted into heat.
For the sake of simplicity, let us assume that all the heat developed is contained in only one body, while the other retains the same temperature and volume. 
Then the entropy value $(S'-S)$ only applies to one body and generally depends not only on the amount of heat developed and the initial temperature $T$ of the body, but also on its temperature and volume changes. 
Only in the case that the amount of heat developed is infinitely small $(=dQ)$, and the possible change in volume has taken place in a reversible manner, that we can set  
\vspace*{-3mm}  
\begin{align}
dS \; = \; \frac{dQ}{T}
\label{Eq_55} \: 
\end{align}
an essentially positive expression, from which it follows that friction, like heat conduction, always forms a natural {\it\color{red}(privileged)} process, which therefore does not form without another process, which therefore cannot be reversed without other compensation
(because if the change in volume does not take place in a reversible manner, i.e. if it forms a natural {\it\color{red}(privileged)} process in itself, then the entropy value of the entire process can only become greater).

Since all real natural {\it\color{red}(privileged)} processes are connected with heat conduction or friction or impact, it can be concluded that neutral {\it\color{red}(indifferent)} processes are only ideal and never occur in reality.
In all its processes, therefore, nature progresses towards a certain goal, and this goal is to increase the {\it\color{red} sum of the entropies} of all existing bodies.
If we ask about the necessary conditions of the state that would correspond to the maximum of {\it\color{red} this sum}, i.e. the absolute equilibrium state of the world, then it is characterized by the fact that all bodies have the same temperatures and exert the same pressure towards the outside, and furthermore that all mechanical work reserves and all mechanical work contributes nothing to increasing the entropy, while the increase in energy generally also increases the entropy.
One can therefore also say that nature generally strives to convert mechanical work into energy, but only to the extent that this increases entropy: the greater the entropy value of a change of state, the easier it is to carry out in nature, and vice versa.

\vspace*{-1mm} 
\begin{center}
-------- \underline{Conclusions} (p.42-43) -------- 
\end{center}
\vspace*{-2mm}

If we now turn again from the special field of heat to the more general subject considered at the beginning of this section, and compare the final proposition expressed there with our present results, we find that the function $(S'-S)$ there, 
which by its sign decides on the possibility of a transition in nature from the first to the second state and vice versa, 
corresponds here to the \dashuline{\,entropy-{\it\color{red}(change)} value $\sum \left( S' \, - \, S \right)$}, which through its value (and like the general function $S$) forms the measure of nature's preference for the corresponding state, and in the special case of heat phenomena is represented by $\sum S$, i.e. the sum of the entropies of all bodies that are included in the state under consideration.
We must therefore regard the entropy value of a change of state, as we have considered it so far, as a special form of \dashuline{\,a more general expression}, which occurs when the changes extend only to the area of heat and mechanical forces.

If we therefore extend the name \dashuline{\,entropy-{\it\color{red}(change)} value} to that \dashuline{\,more general function}, then the \dashuline{\it theorem of entropy value now applies to all occurring natural {\it\color{red}(privileged)}  phenomena} and \dashuline{{\it the second law}} of the mechanical heat theory \dashuline{{\it then appears}}, like the first, \dashuline{{\it as the consequence of a more general natural}}
\dashuline{{\it  {\it\color{red}(privileged)} principle}}.

Of course, this generalization is initially only hypothetical and must first be checked for accuracy through closer investigations.

Even in the area of heat, the proof is not yet binding for all bodies without exception because, as you can see, the statements made in this section only apply to bodies whose state is completely determined by temperature and volume and which exert a positive, normal pressure on the external environment that is the same on all sides.
The generalization to all bodies would initially be associated with considerable inconvenience, but probably not with serious difficulties.

\vspace*{-1mm} 
\begin{center}
--------------------------------------------- 
\end{center}
\vspace*{-4mm}



\section{\underline{Second Section {\color{red}of the Thesis}} (p.44-61)}
\vspace*{-2mm}

The following statements are based on \dashuline{\,the principle first expressed by R. Clausius\,} in his famous treatises on mechanical heat theory:
\vspace*{1mm} \\ \hspace*{1mm} 
{\bf that in nature a heat transfer from a colder to a warmer body cannot take place without compensation.}

Although \dashuline{\,this theorem\,} initially met with much opposition from the most important authorities, it was able to be successfully defended in every single case by its discoverer, so that it can now be found cited as a theorem in the textbooks on the mechanical theory of heat.

However, in the above version, \dashuline{\,this sentence still appears somewhat vague and ambiguous}, as long as a mathematical expression does not provide a very specific measure for the compensation required in each individual case.
And this circumstance, that the measure of compensation was variously and incorrectly understood, perhaps explains the fact that the sentence itself had to experience so many challenges.
Because it is clear that, depending on the value of this compensation, the sentence as it stands in its above form can be interpreted in different ways and thus leads to very different consequences, as we will clearly see in the following.

In the previous section we derived directly from the second law what is to be regarded as the generally valid measure for the compensation required for such a heat transfer.
However, we would like to focus here on the consequences that arise from the application of the above principle, if one generally assumes the equivalent value of this transformation for the amount of compensation required in the transformation of heat from a lower to a higher temperature, as first stated by Clausius.

In his latest edition of the mechanical theory of heat (1876), Clausius treats only two types of transformations, namely the transformation of a quantity of heat from one temperature to another and the transformation of heat into work, and vice versa.
We therefore only want to consider these two types of transformation here.
Each of the two is assigned a certain \dashuline{\,equivalence values}, which is known: \vspace*{1mm} \\
1) for the transformation of the amount of heat $Q$ (measured according to mechanical mass) from the (absolute) temperature $T_1$ into the temperature $T_2$:
\vspace*{-1mm}  
\begin{align}
Q \:.\; 
\left( \frac{1}{T_2} \:-\: \frac{1}{T_1} \right)
\label{Eq_56} \: ;
\end{align}
2) for the transformation of the amount of heat $Q$ from the temperature $T$ into work: 
\vspace*{-1mm}  
\begin{align}
- \: \frac{Q}{T}
\label{Eq_57} \: ;
\end{align}
and vice versa for the transformation of work $Q$ into heat at temperature $T$: 
\vspace*{-1mm}  
\begin{align}
\: \frac{Q}{T}
\label{Eq_58} \: .
\end{align}
It might perhaps be useful to briefly outline the results of the investigations carried out under this assumption: the application of Clausius' principle, if we assume the above \dashuline{\,equivalence values\,} as the \dashuline{\,general measure of the compensation} required therein, will lead us to a number of conclusions which are in such harmony with the previous results of the mechanical heat theory that we must regard them as absolutely correct.
On the other hand, however, we will recognize that the sentence obtained in this way in no way leads to the second law, or is itself somehow related to it, but that its meaning lies on a completely different side.

Now let's move on to executing the developed plan.
In order to distinguish the terms precisely, we want to apply the term ``\,transformation\,'' only to the transfer of a single amount of heat from one temperature to another, or from heat to work and vice versa; then we can say: any process in nature that brings about changes in heat and work generally includes a series of transformations, each of which has its specific equivalence value according to the assumption.
We can call the sum of all these equivalence values the \dashuline{equivalence value of the entire process\:}.

So every process has its specific equivalence value, namely the sum of the equivalence values of all associated transformations.
In this sum, the negative terms are caused by the transformation of quantities of heat from lower to higher temperatures, as well as from heat into work. See (\ref{Eq_56}) and (\ref{Eq_57})

If we now regard the (negative) equivalent value of this transformation as the measure for the compensation required in Clausius' principle when heat is transformed from a lower to a higher temperature, then this means that this transformation is different because of its possibility in nature transformation with a positive equivalent value, which is at least as large as the absolute value of the equivalent value of the first transformation. That would be a transformation of heat from a higher to a lower temperature, or of work into heat.
Compare (\ref{Eq_56}) and (\ref{Eq_58}). It then follows that every transformation of heat into work also requires compensation because its equivalent value is also negative, meaning that it is equivalent to a transformation of heat from a lower to a higher temperature.
Now since every process, as is natural, represents a possible change in nature, it follows that in the sum, which forms its equivalent value, the negative members must be compensated by the positive members in such a way that their absolute value is in any case not greater than the latter, which can be expressed in other words
\vspace*{-3mm} \begin{quotation}
\hspace*{15mm}
\dashuline{(Theorem:)\:}
{\bf the \dashuline{\:equivalent value} of a process cannot be negative\,.}
\end{quotation}
\vspace*{-2mm}
The \dashuline{\,following two sentences\,} will result from  \dashuline{\:this theorem\,} as strict consequences: 
\vspace*{1mm} \\
{\bf 1) the \dashuline{equivalent value} of every process is $=0\:$;}
\vspace*{1mm} \\
{\bf 2) the true {\it\color{red}(specific)} heat capacity of every body is constant, i.e. independent of temperature and volume\,.}

The intimate connection between these two theorems will be discussed in detail below, after we have derived them from the above premise.
At the same time, however, it can already be seen here that they are in no way related to the second main theorem of the mechanical theory of heat, simply because they make no distinction at all between reversible and non-reversible processes.

We derive these two theorems from the premise that the equivalent value of a process cannot be negative, first by considering reversible processes.

\subsection{\underline{Reversible processes} (p.47-49)}
\vspace*{-1mm}

It follows immediately that the equivalent value of any reversible process is $=0$, because according to the premise it could also be positive at first, but then the reversed process would have a process with a negative equivalent value, because in this process the equivalent values of all individual transformations have the opposite sign. This case also contradicts the premise, so the equivalent value of a reversible process can only be $=0$.

Let us apply this result to the following reversible process: Any body whose state is determined by temperature and volume is compressed in a reversible manner by an external pressure, i.e. so that the external pressure differs imperceptibly little from the counter-pressure of the body. No heat should be absorbed from outside or released to the outside.

Let us now consider at some point in the process an elementary part of it, which of course also forms a reversible process in itself.
If $dI$ denotes the internal work done during this elementary process, $dW$ the external work done, $dH$ the increase in heat of the body, and finally $T$ and $T+dT$ the temperatures at the beginning and end of the process, then we have to set:
\vspace*{-1mm}  
\begin{align}
0 \;=\; dH \;+\; dI \;+\; dW 
\label{Eq_59} \: .
\end{align}

According to the above, the \dashuline{equivalent value} of this elementary process is $=0$.
This elementary process consists of the following elementary transformations, whose equivalence values we will add immediately, neglecting the infinitely small second-order quantities:\vspace*{-2mm}
\begin{enumerate}
\item
\!\!\!) The work $-\:dW$ (which is the absolute value in the case of compression) is converted into heat of temperature $T$, with the \dashuline{equivalence value} 
\vspace*{-2mm}
$$-\,\frac{dW}{T} \: ;$$
\vspace*{-4mm}
\item
\!\!\!) The work $-\:dI$ is converted into heat of temperature $T$, with the \dashuline{equivalence value} 
$$-\,\frac{dI}{T}  \: ;$$
\vspace*{-4mm}
\item
\!\!\!) The heat $H$ of the body is transformed from the temperature $T$ into the temperature $T+dT$, with the \dashuline{equivalence value} 
\vspace*{-2mm}
$$ H\:.\:\left( \frac{1}{T\:+\:dT}\:-\:\frac{1}{T} \right)
   \; \approx \; - \,\frac{H\:.\;dT}{T^2} \: .$$
\end{enumerate}

By adding these three expressions we get the \dashuline{\,equivalent value of the elementary process}: 
\vspace*{-1mm}  
\begin{align}
   -\,\frac{dW}{T}
\; -\,\frac{dI}{T} 
\; -\,\frac{H\:.\;dT}{T^2} 
\;=\; 0
\label{Eq_60} \: ,
\end{align}
but from (\ref{Eq_59}) we get: 
$$-\:dW \;-\; dI  \;=\; dH $$
which can be substituted into (\ref{Eq_60}) to give: 
\vspace*{-1mm}  
\begin{align}
\frac{dH}{T} 
\; -\,\frac{H\:.\;dT}{T^2} 
\;=\; 0
\nonumber \: ,
\end{align}
or equivalently: 
\vspace*{-3mm}  
\begin{align}
\frac{dH}{H} \;=\; \frac{dT}{T} 
\nonumber \: ,
\end{align}
and hence: 
\vspace*{-3mm}  
\begin{align}
H \;=\; c \:.\: T
\label{Eq_61} \: ,
\end{align}
where \dashuline{\:$c$ is a constant that only depends on the type of substance and its mass\,}, but not on volume and temperature, because $dH$ represents the complete differential of the quantity $H$ with respect to volume and temperature. 

So the heat of a body is proportional to the absolute temperature and a quantity $c$ that remains constant for all changes of state, i.e: \vspace*{-3mm}
\begin{quotation}
\hspace*{15mm}
{\bf The true {\it\color{red}(specific)} heat capacity of every body is constant.}
\end{quotation}


\subsection{\underline{Irreversible processes} (p.49-61)}
\vspace*{-1mm}

It is now unnecessary to provide the general proof that the \dashuline{\,equivalence value of every process is $=0$},
because this can be done directly with the results obtained.

The means by which the state of one or more bodies can be altered with respect to volume and temperature consist in the supply of heat (by conduction or radiation) and in the exertion of pressure, if we confine ourselves, as we always do here, to the field of mechanical forces and heat.
Therefore, we first want to investigate a process that occurs with two bodies in which they exchange heat, but at the same time are exposed to any pressure forces and can change their volumes.

The true heat capacities of the two bodies, based on their masses, are $c_1$ and $c_2$.
Let us now consider again an elementary process, at the beginning of which the temperatures of the two bodies are $T_1$ and $T_2$, and at the end of which they may be $T_1+dT_1$ and $T_2+dT_2$.
If during this elementary process the amount of heat $dQ$ (which can also be negative or $=0$) has passed from the first body to the second, we have the equations:
\vspace*{-3mm}  
\begin{align}
\mbox{For the first body:}\;\;\;\;
-\,dQ \;=\; c_1 \: dT_1 \;+\; dI_1 \;+\; dW_1
\label{Eq_62} \: , \\
\mbox{For the second body:}\;\;\;\;
+\,dQ \;=\; c_2 \: dT_2 \;+\; dI_2 \;+\; dW_2
\label{Eq_63} \: , \\
\mbox{Hence:}\;\;\;\;
0 \;=\; 
c_1 \: dT_1 \;+\; dI_1 \;+\; dW_1
\;+\;
c_2 \: dT_2 \;+\; dI_2 \;+\; dW_2
\label{Eq_64} \: .
\end{align}
Here $dI$ denotes the internal work and $dW$ the external work, including any living force {\it\color{red} (kinetic energy)} generated if the external pressure and the counter-pressure of a body are of different magnitudes.
The elementary process includes the following transformations, whereby we neglect the infinitely small second-order quantities:
\vspace*{-3mm}
\begin{enumerate}
    \item 
\!\!\!) The {\it\color{red}sum of} work and living force {\it\color{red} (kinetic energy)} $-(dI_1+dW_1)$ has turned into heat of temperature $T_1$, with an \dashuline{\,equivalence value}
             $$-\frac{dI_1\:+\:dW_1}{T_1} \; ,$$ 
by setting the equivalent value of the transformation of living force {\it\color{red} (kinetic energy)} into heat equal to that of the transformation of work into heat, which can be implemented as dictated by the circumstances.
    \item 
\!\!\!) The heat of the first body $c_1\:T_1$ is brought from the temperature $T_1$ to the temperature $T_1+dT_1$, with the \dashuline{\,equivalence value} 
   $$c_1\;dT_1 \;.\:
   \left( \frac{1}{T_1\:+\:dT_1}\:-\:\frac{1}{T_1}  \right)
    \; \approx \;
    -\frac{c_1\:.\:dT_1}{T_1} \; .$$
    \item 
\!\!\!) The amount of heat $dQ$ has been transformed from the temperature $T_1$ partly into heat from the temperature $T_2$, partly into internal and external work and living force {\it\color{red} (kinetic energy)}, according to equation (\ref{Eq_63}):  
\\ \hspace*{3mm}
a) the part $c_2\:dT_2$ is converted into heat of temperature $T_2$, with the \dashuline{\,equivalence value} 
   $$c_2\;dT_2 \;.\:
   \left( \frac{1}{T_2}\:-\:\frac{1}{T_1}  \right) \; ,$$
\\ \hspace*{3mm}
b) the part $(dI_2+dW_2)$ transformed into work and living power  {\it\color{red} (kinetic energy)}, 
\\ \hspace*{9mm}with the \dashuline{\,equivalence value} 
         $$-\frac{dI_2\:+\:dW_2}{T_1} \; .$$ 
    \item 
\!\!\!) The heat of the second body $c_2\:dT_2$ is transformed from the temperature $T_2$ into the temperature $T_2+dT_2$, with the \dashuline{\,equivalence value} 
   $$c_2\;dT_2 \;.\:
   \left( \frac{1}{T_2\:+\:dT_2}\:-\:\frac{1}{T_2} \right)
    \; \approx \;
    -\frac{c_2\:.\:dT_2}{T_2} \; .$$
\end{enumerate}
Therefore, by adding these $5$ equivalence values as the \dashuline{\,equivalence value of the elementary process\,}, we get the expression: 
$$
\; - \: \frac{dI_1\:+\:dW_1}{T_1} 
\; - \: \frac{c_1\:.\:dT_1}{T_1} 
\; + \: c_2\;dT_2 \:.
   \left( \frac{1}{T_2}\:-\:\frac{1}{T_1}  \right) 
\; - \: \frac{dI_2\:+\:dW_2}{T_1} 
\; - \: \frac{c_2\:.\:dT_2}{T_2} \; ,
$$ 
otherwise written as: 
$$ -\:
\frac{\left( \: c_1 \: dT_1 \;+\; dI_1 \;+\; dW_1
\;+\;
c_2 \: dT_2 \;+\; dI_2 \;+\; dW_2 \: \right)}{T_1} \; , $$ 
which according to equation (\ref{Eq_64}) is $=0$.

Since the equivalence value of the entire process is obviously equal to the sum of the equivalence values of all elementary processes, \dashuline{the entire process has the equivalence value $0$\,}.

The most general case of a process is that a series of bodies undergo changes in volume and temperature through the mutual communication of heat and the exertion of forces on one another, whereby we consider all bodies which take part in these changes to be included in the process.
In order to determine the equivalence value of such a process, one breaks it down into a series of elementary processes and adds their equivalence values. If one of these elementary processes imparts changes to only two bodies, as in the case previously considered, then its equivalence value is proven to be $=0$. But if it is of a more complicated nature, in that several bodies act on one another at the same time, the problem can be solved by decomposition, in that the elementary process can be thought of as carried out according to the law of superimposition of small effects in such a way that only two bodies are subject to changes at the same time.

But then we have the previous case, so the equivalence value of each elementary process and also that of the entire process is $=0$. 
\dashuline{\,We can therefore state the sentence:\,} 
\vspace*{-1mm} \begin{quotation}
\hspace*{15mm}
{\bf The equivalence value of every process is $=0$.}
\end{quotation}

The mathematical expression of this sentence follows immediately: if a process consists of the following transformations: 
\vspace*{1mm} \\ \hspace*{15mm}
an amount of heat $Q_1$ from temperature $T_1$ into $T'_1$ 
\vspace*{1mm} \\ \hspace*{15mm}
an amount of heat $Q_2$ from temperature $T_2$ into $T'_2$ 
\vspace*{1mm} \\ \hspace*{15mm}
an amount of heat $Q_3$ from temperature $T_3$ into $T'_3$ 
\vspace*{1mm} \\ \hspace*{15mm}
... ... ... ... ... 
... ... ... ... ... 
... ... ... ... ... 
... ... ... 
\vspace*{1mm} \\
then we have: 
$$
Q_1 \:.\: \left(\frac{1}{T'_1}\:-\:\frac{1}{T_1}\right)\;+\;
Q_2 \:.\: \left(\frac{1}{T'_2}\:-\:\frac{1}{T_2}\right)\;+\;
Q_3 \:.\: \left(\frac{1}{T'_3}\:-\:\frac{1}{T_3}\right)\;+\;
. . . .
\;=\; 0$$

This also includes the transformations of work and living force {\it\color{red} (kinetic energy)} into heat and vice versa, if one understands work or living force {\it\color{red} (kinetic energy)} 
as a quantity of heat $Q$ of temperature $\infty$.

This abbreviation, although it does not correspond to any real fact, is very suitable for putting the formulas into a more convenient form.
In fact, the transformation of heat $Q$ from temperature $T$ into work has the equivalent value: 
$$ Q \;.\, \left( \frac{1}{\infty} \:-\: \frac{1}{T} \right)
\; = \; -\: \frac{Q}{T} \: , $$ 
while the reverse transformation has the equivalent value 
$$ Q \;.\, \left( \frac{1}{T} \:-\: \frac{1}{\infty} \right)
\; = \; +\: \frac{Q}{T} \: . $$ 
Using this abbreviation, \dashuline{the theorem can be written as follows\,}:  
\vspace*{-1mm}  
\begin{align}
\boxed{\:
\sum Q \;.\, \left( \frac{1}{T'} \:-\: \frac{1}{T} \right)
\; = \; 0 \:}
\label{Eq_65} \; ,
\end{align}
\dashuline{which is valid for every process\,}, where $Q$ is the general expression of a quantity of heat which has the temperature $T$ at the beginning of the process and the temperature $T'$ at the end of the process

We now want to make a few \dashuline{\,applications of this theorem\,} to simple processes, and we will see how it proves to be true in each individual case if we only take into account all the transformations that occur.


\vspace*{-1mm} 
\begin{center}
-------- \underline{First example (p.53-54)} -------- 
\end{center}
\vspace*{-2mm}

If we let a \dashuline{\,quantum of perfect gas} with the true {\it\color{red}(specific)} heat capacity $c$ (based on the mass of the gas) expand under work in a heat-impermeable shell, whereby the temperature drops from $T$ to $T'$, we have:
\vspace*{1mm} \\
1) the negative transformation of heat into work,
\vspace*{1mm} \\
2) the positive transformation of the remaining heat of the gas from a higher to a lower temperature.

It can be seen directly that \dashuline{the corresponding equivalence values cancel each other out\,}. 
Indeed, the work done is of the same magnitude as the amount of heat lost, i.e. $=c\:.\:(T-T')$.
The equivalent value of the transformation from temperature $T$ into work is therefore 
$$ -\:\frac{c\:.\:(T-T')}{T} \; , $$
and the equivalent value of the transformation of the remaining heat $c\:.\:T'$ from the temperature $T$ into the temperature $T'$ is 
$$ +\:c\:.\:T' \:.\: 
\left( \frac{1}{T'} \:-\: \frac{1}{T} \right) \; , $$
and finally, by addition we get 
$$ -\:\frac{c\:.\:(T-T')}{T}
 \;+\:c\:.\:T' \:.\: 
\left( \frac{1}{T'} \:-\: \frac{1}{T} \right)
\;=\; 0 \; ,$$ 
\dashuline{\,as it should be\,}.


\vspace*{-1mm} 
\begin{center}
-------- \underline{Second example (p.54-58)} -------- 
\end{center}
\vspace*{-2mm}

We imagine two perfect gases in arbitrary quantities, with the true {\it\color{red}(specific)} heat capacities $c_1$ and $c_2$ (based on the masses) and the temperatures $T_1$ and $T_2$, where $T_1>T_2$.

We want to \dashuline{calculate the equivalent value of the process} that is carried out when the amount of heat $Q$ passes from the first to the second gas while the volumes remain constant.
If we denote the final temperatures of the two gases by $T'_1$ and $T'_2$, we have:
\vspace*{-2mm}  
\begin{align}
\mbox{for the first gas:}\;\;\;\;
c_1 \: T_1 \;-\; Q \;=\;  c_1 \: T'_1
\label{Eq_66} \: , \\
\mbox{for the second gas:}\;\;\;\;
c_2 \: T_2 \;+\; Q \;=\;  c_2 \: T'_2
\label{Eq_67} \: .
\end{align}
The process consists of $3$ transformations:
\begin{enumerate} 
  \item \!\!\!) 
the amount of heat $Q$ is transformed from the temperature $T_1$ into the temperature $T'_2$, with the equivalence value: 
$$ Q \;.\,
\left( \frac{1}{T'_2} \:-\: \frac{1}{T_1} \right)\; ;$$
  \item \!\!\!) 
the heat $c_1\:T'_1$ remaining in the first gas is transformed from the temperature $T_1$ into the temperature $T'_1$, with the equivalence value: 
$$ c_1\:.\:T'_1 \;. 
\left( \frac{1}{T'_1} \:-\: \frac{1}{T_1} \right)
\;=\; c_1 \:-\: c_1 \: \frac{T'_1}{T_1} 
\;,\;\;\mbox{and according to (\ref{Eq_66})}\,
= \; \frac{Q}{T_1} \: ;$$
  \item \!\!\!) 
the heat $c_2\:T_2$ originally present in the second gas is transformed from the temperature $T_2$ into the temperature $T'_2$, with the equivalence value: 
$$ c_2\:.\:T_2 \;. 
\left( \frac{1}{T'_2} \:-\: \frac{1}{T_2} \right)
\;=\; \frac{c_2 \: T_2}{T'_2} \:-\: c_2  
\;,\;\;\mbox{and according to (\ref{Eq_67})}\,
= \; -\,\frac{Q}{T'_2} \: .$$
\end{enumerate} 

By adding these three expressions, we get \dashuline{\:the equivalent value of the process\,}, which, as we can immediately see, \dashuline{\;is $=0$, as the theorem requires\,}.

What is noteworthy here is that this result holds no matter how large $Q$ may be, and even if $Q$ is negative, i.e. when heat transfers from the colder gas to the warmer one, whereby of course we can no longer speak of a process, but only of an imaginary change of state.
In this case, the negative transformations of the transferred heat quantity $Q$ (and the heat originally present in the first gas) from a lower to a higher temperature are cancelled out by the positive transformation of the heat remaining in the second gas from a higher to a lower temperature.

Furthermore, the theorem applies unchanged, even if the dimensions of the two gases are assumed to be so large (in comparison to the amount of heat $Q$ transferred) that the temperatures $T_1$ and $T_2$ of both gases are only imperceptibly changed by the heat transfer, and can be regarded as virtually constant.
Nevertheless, in this case the equivalent value of the process in question retains the value $0$, because the transformation of the quantity of heat $Q$ from the temperature $T_1$ into the temperature $T_2$ is cancelled out by the transformations of the quantities of heat contained in the two gases, whose equivalent values retain finite values because in them the unlimited smallness of the one factor is balanced by the unlimited size of the other, namely the amount of heat transformed.

In a similar way, the validity of the theorem of the equivalence value of a process can be traced to other phenomena in the field of heat, e.g. friction and the motion of a body in a resisting medium.
Here we have the positive transformation of work or living force {\it\color{red} (kinetic energy)} into heat, and also the negative transformation of the heat already present in the bodies from a lower to a higher temperature. 
\dashuline{According to the theorem}, \dashuline{\,both equivalence values cancel each other out\,}.

We have, of course, derived the proposition in its general form from a premise which we cannot regard as necessary 
at first, 
namely by introducing the equivalent values of the transformations without further justification as the measure of the compensation required for the transformation of heat from a lower to a higher temperature and of heat into work. 
But we are nevertheless led to recognize this theorem (about the equivalence value of a process) as generally valid, because of the close connection in which it stands with the theorem about the constancy of the true {\it\color{red}(specific)} heat capacity of a body, since the latter has already been shown to be highly probable through other investigations.

The fact that the two sentences are mutually dependent can already be seen from the above argument.
But we can also see directly that they are virtually identical.
If we write the theorem of the equivalent value of a process according to (\ref{Eq_65}) 
\begin{align}
\sum \: Q \:
\left( \frac{1}{T'} \:-\: \frac{1}{T} \right)
\; = \; 0
\nonumber \: 
\end{align}
in the form 
\vspace*{-5mm}
\begin{align}
\boxed{\:
\sum \: \frac{Q}{T'}  
\;\; = \;\; 
\sum \: \frac{Q}{T}  
\:}
\label{Eq_68} \; ,
\end{align}
where the value of the right-hand side of the equation is given by the initial state, that of the left-hand side by the final state of the process, and one has to sum up the amounts of heat present for each state, each divided by its temperature.
However, the terms corresponding to existing quantities of work or living forces {\it\color{red} (kinetic energy)} are omitted because they have $\infty$ in the denominator, and thus \dashuline{\,the theorem\,} expresses that the sum of the real quantities of heat, each divided by its temperature, is not changed by the process, which is why \dashuline{\,we can write\,}:
\begin{align}
\sum \: \frac{Q}{T}  
\; = \;  const.
\label{Eq_69} \; 
\end{align}
Now, $Q/T$ is nothing other than the product of the mass of the body in question and its true heat capacity. 
If we therefore apply the \dashuline{\,theorem} to a process that takes place with a single body, the equation
$$\frac{Q}{T} \;=\; const. $$ 
immediately gives us the \dashuline{\,theorem of the constancy of the true {\it\color{red}(specific)} heat capacity\,}. 
From this the identity of the two sentences discussed becomes clear.

However, the \dashuline{\,principle of the equivalence of a process\,} has a somewhat more general form in that it can be extended to other natural {\it\color{red}(privileged)} processes, e.g. on chemical processes.

Let us consider as the initial state of a chemical process various separate substances that can chemically combine with each other, with the masses $m_1, \;m_2, \;...$, and the true {\it\color{red}(specific)} heat capacities $c_1, \;c_2, \;...$, and as the final state their chemical product with the mass $M$ and the true {\it\color{red}(specific)} heat capacity $C$, then equation (\ref{Eq_69}) $(\sum Q/T = const.)$ gives us the following relation 
\begin{align}
\sum \: m \; c 
& \: = \; M \: . \: C  
\nonumber \\
\mbox{or}\;\;\;\;
C \: . \: \sum \: m
& \: = \; \sum \: m \: c
\label{Eq_70} \; ,
\end{align}
i.e. the product of the weight and the true {\it\color{red}(specific)} heat capacity of a body is equal to the sum of the products of the weights and true {\it\color{red}(specific)} heat capacities of its chemical constituents.


\vspace*{-1mm} 
\begin{center}
-------- {\underline{{\it\color{red}General Conclusions (p.58-61)}}} -------- 
\end{center}
\vspace*{-2mm}

It is clear from all of the above that it was our intention to show that the \dashuline{\,equivalent values\,} of the transformations discussed in this section cannot form the actual measure of the compensation required by Clausius' principle for a heat transfer from a colder to a warmer body.
For although its introduction and application \dashuline{\,led to a theorem whose correctness leaves no room for doubt\,}, the meaning of this theorem lies in a completely different area  \dashuline{\,and refers to a completely different law of nature\,} than that which Clausius' principle is intended to express.

While the latter has the task of establishing a constant progress in the processes of nature, in such a way that certain processes that can be carried out in one direction cannot be carried out in the opposite direction 
(which documents its relationship to the second law),
this actual core of Clausius' principle does not find its corresponding expression in the sentence of the \dashuline{\,equivalence value of a process\,}, 
for the reason that in this \dashuline{\,theorem\,} the compensation necessary for a heat transfer from a colder to a warmer body is essentially different from that in the {\it(Clausius')} principle.
Because, in the \dashuline{\,theorem\,} of the \dashuline{\,equivalent value of a process\:}, this compensation already lies in the fact that the two bodies between which the heat transfer takes place undergo temperature changes through the same process, and consequently the quantities of heat they contain undergo transformations.
The latter transformations precisely compensate for the transformation of the amount of heat transferred, as we saw above in the second example in a special case.
 But this type of compensation takes place with every heat transfer, even from a warmer to a colder body, indeed with every process in general, and, as we show, only requires the fulfilment of the sole condition that the true {\it\color{red}(specific)} heat capacity of every body is constant.
It is therefore easy to see how fundamentally different this conception of compensation, to which the introduction of the \dashuline{\,equivalence values\,} discussed necessarily leads, is from that which is essentially in the sense of \dashuline{\,Clausius' principle\,}, since the latter is not responsible for the transformation of an \dashuline{\,individual amount of heat\,}, but rather requires compensation for an \dashuline{\,entire change in state\,}.
 We must therefore deny the \dashuline{\,equivalence values of the individual transformations\,} the general significance that they are supposed to have for Clausius' principle, and thus also for the second law.

Finally, there is the opportunity here to refute an objection that might be raised against our comments.

As is well known, the second law in its application to a circular-{\it\color{red}(cyclic)} process carried out by any body, as first derived by Clausius and as we developed it in the previous section, reads as follows
\begin{align}
- \,\int \frac{dQ}{T}  \; \geq \; 0 \; 
\nonumber
\end{align}
according to equation (\ref{Eq_34}), where $dQ$ denotes the amount of heat absorbed by the body at any given time and $T$ its temperature.
Now, as is well known, Clausius obtained the expression $-\int dQ/T$ by adding the \dashuline{\,equivalent values\,} of the transformations of all the quantities of heat $dQ$ from their respective temperatures $T$ into work, and one could therefore 
claim 
that in this case the sum of the \dashuline{\,equivalent values\,} of all the transformations that have occurred, i.e. the \dashuline{\,equivalent value\,} of the circular-{\it\color{red}(cyclic)} process, does not necessarily have the value $0$, as required by the \dashuline{\,theorem of the equivalent value\,} of a process, but rather that this \dashuline{\,equivalent value\,}, depending on whether it is $=0$ or positive, decides whether the circular-{\it\color{red}(cyclic)} process can be reversed or not, and that it is therefore of essential importance for the second law.

The answer to this is that the \dashuline{\,theorem of equivalence\,} of a process applies to this circular-{\it\color{red}(cyclic)} process, as it does to every process. But the expression $-\int \, dQ/T$ does not represent the \dashuline{\,equivalence value\,} of the cycle, because it does not include the \dashuline{\,equivalence values\,} of all transformations that have occurred.
This is because it does not take into account all the transformations that the quantities of heat in the heat reservoirs undergo as a result of the temperatures of the reservoirs changing due to the release or absorption of heat.
The \dashuline{\,equivalent values\,} of these transformations must by no means be neglected, because even if the dimensions of the heat reservoirs are assumed to be so large that the changes which their temperatures undergo in the course of the process are almost negligible, this may only be done on condition that the quantities of heat contained in the reservoirs are assumed to be large over all dimensions.
But since the \dashuline{\,equivalent value\,} of a transformation has the transformed quantity of heat as a factor, its value cannot be made to disappear by the above assumption.

If the last transformations discussed were also taken into account, the sum of the \dashuline{\,equivalent values\,} of all transformations that occurred would be the value $0$, as we have generally proven.
To make matters worse, we also want to prove this fact directly here.  
If, in order to avoid transformations of work quantities, we assume only perfect gases as heat reservoirs, then the amount of heat contained in such a gas container is $c \: T$, where $c$ represents the true {\it\color{red}(specific)} heat capacity of the gas, based on its mass.
Due to the loss of the amount of heat $dQ$, the temperature should fall to $T-dT$, so that if the volume remains constant, as in equation (\ref{Eq_33}): $dQ = c \:.\:dT$.

Then the \dashuline{\,equivalent value\,} of the transformation of the amount of heat ``\,$c\:T$\:'' of the temperature $T$ into the temperature $T-dT$ is: 
\begin{align}
c\:T\:.\:
\left(\frac{1}{T\:-\:dT}\;-\;\frac{1}{T}\right)
& \: \approx \; \frac{c\:.\:dT}{T}
\; , \:\:\:\mbox{and according to (\ref{Eq_33}):}\:
\;=\; \frac{dQ}{T}
\nonumber \; .
\end{align}

If one integrates over all reservoirs, one obtains $\boxed{\:\int dQ/T\:}$ as the sum of the \dashuline{equivalence values\,} of the corresponding transformations, and this sum added to the expression $\boxed{\:-\,\int dQ/T\:}$ gives the value $0$ as the \dashuline{\,equivalence value\,} of the circular-{\it\color{red}(cyclic)} process, as it must be.

Now that it has been shown that the sum $-\,\int dQ/T$ does not include all the transformations included in the circular-{\it\color{red}(cyclic)} process, the assertion that the \dashuline{equivalence values of the transformations\,} are of essential importance for the second law is also invalid.
For if one wants to introduce them once, one must in any case also take into account all the transformations that occur, and not only a part of them, or one would at least have to establish a striking difference between those that are taken into account and those that are not.
However, this has not yet been done, and it is therefore highly recommended to define the expression $\boxed{\:-\,\int dQ/T\:}$ not as a sum of \dashuline{equivalence values\,}, but as the \dashuline{entropy-{\it\color{red}(change)} value of the circular-{\it\color{red}(cyclic)} process\,}, as we did in the first section. 


\vspace*{-1mm} 
\begin{center}
--------------------------------------------------- 
\end{center}
\vspace*{-2mm}


\vspace*{2mm} 
{\it ``\,Planck, Second Law\,''}

\vspace*{2mm} 
{\it ``\,Royal Court. Printed by E. Mühlthaler in Munich\,''}

\end{document}